\documentclass[3p,times,procedia]{elsarticle}

\usepackage{subfigure}  

\usepackage{amssymb}

\usepackage[figuresright]{rotating}

\newcommand{\up}[1]{$^{#1}$}

\newcommand{\cst}[2]{\mbox{#2$\times$10$^{-#1}$\,cm$^2$}}

\newcommand{\powero}[1]{\mbox{10$^{#1}$}}
\newcommand{\powert}[2]{\mbox{#2$\times$10$^{#1}$}}
\newcommand{\ru}{\mbox{events/(kg$\cdot$d)}}
\newcommand{\dru}{\mbox{events/(keV$_{ee}$$\cdot$kg$\cdot$d)}}

\newcommand{\um}{$\mu$\mbox{m}}

\newcommand{\pbten}{$^{210}$Pb}
\newcommand{\biten}{$^{210}$Bi}
\newcommand{\ptwo}{\mbox{$^{32}$P}}
\newcommand{\sitwo}{\mbox{$^{32}$Si}}
\newcommand{\indium}{\mbox{$^{115}$In}}
\newcommand{\ber}{\mbox{$^{7}$Be}}
\newcommand{\lis}{\mbox{$^{7}$Li}}

\newcommand{\sodium}{\mbox{$^{22}$Na}}
\newcommand{\ironfive}{\mbox{$^{55}$Fe}}

\newcommand{\rad}{\mbox{$^{226}$Ra}}
\newcommand{\tho}{\mbox{$^{232}$Th}}
\newcommand{\urafive}{\mbox{$^{235}$U}}
\newcommand{\urafour}{\mbox{$^{234}$U}}
\newcommand{\ura}{\mbox{$^{238}$U}}
\newcommand{\amer}{\mbox{$^{241}$Am}}
\newcommand{\cali}{\mbox{$^{252}$Cf}}

\newcommand{\gev}{\mbox{GeV/}$c^2$}
\newcommand{\dedx}{$dE/dx$}
\newcommand{\evr}{\mbox{eV$_{\rm r}$}}
\newcommand{\eve}{\mbox{eV$_{ee}$}}
\newcommand{\swn}{\mbox{$\sigma_{\chi \mbox{\sc n}}$}}

\begin{document}

\begin{frontmatter}




\title{DAMIC at SNOLAB}


\author[kicp]{Alvaro~E.~Chavarria}
\author[fnal]{Javier~Tiffenberg}
\author[unam]{Alexis~Aguilar-Arevalo}
\author[mich]{Dan~Amidei}
\author[cam]{Xavier~Bertou}
\author[fnal]{Gustavo~Cancelo}
\author[unam]{Juan~Carlos~D'Olivo}
\author[fnal]{Juan~Estrada}
\author[fnal]{Guillermo~Fernandez~Moroni}
\author[fnal]{Federico~Izraelevitch}
\author[zurich]{Ben~Kilminster}
\author[mich]{Yashmanth~Langisetty}
\author[zurich]{Junhui~Liao}
\author[fiuna]{Jorge~Molina}
\author[kicp]{Paolo~Privitera}
\author[unam]{Carolina~Salazar}
\author[unam]{Youssef~Sarkis}
\author[fnal]{Vic~Scarpine}
\author[mich]{Tom~Schwarz}
\author[cam]{Miguel~Sofo~Haro}
\author[unam]{Frederic~Trillaud}
\author[kicp]{Jing~Zhou}

\address[kicp]{Kavli Institute for Cosmological Physics and The Enrico Fermi Institute, The University of Chicago, Chicago, IL, United States}
\address[fnal]{Fermi National Accelerator Laboratory, Batavia, IL, United States}
\address[unam]{Universidad Nacional Autonoma de Mexico, Mexico D.F., Mexico}
\address[mich]{University of Michigan, Ann Arbor, MI, United States}
\address[cam]{CNEA/CONICET, Centro Atomico Bariloche, San Carlos de Bariloche, Argentina}
\address[zurich]{University of Zurich, Zurich, Switzerland}
\address[fiuna]{Facultad de Ingenieria, Universidad Nacional de Asuncion (FIUNA), Asuncion, Paraguay}

\begin{abstract}
We introduce the fully-depleted charge-coupled device (CCD) as a particle detector. We demonstrate its low energy threshold operation, capable of detecting ionizing energy depositions in a single pixel down to 50\,\eve. We present results of energy calibrations from 0.3\,k\eve\ to 60\,k\eve, showing that the CCD is a fully active detector with uniform energy response throughout the silicon target, good resolution (Fano $\sim$0.16), and remarkable linear response to electron energy depositions. We show the capability of the CCD to localize the depth of particle interactions within the silicon target. We discuss the mode of operation and unique imaging capabilities of the CCD, and how they may be exploited to characterize and suppress backgrounds. We present the first results from the deployment of 250\,\um\ thick CCDs in SNOLAB, a prototype for the upcoming DAMIC100. DAMIC100 will have a target mass of 0.1\,kg and should be able to directly test the CDMS-Si signal within a year of operation.
\end{abstract}


\end{frontmatter}


\section{DAMIC: Dark Matter in CCDs}
\label{sec:intro}

There is strong cosmological and astrophysical evidence supporting the existence of non-baryonic, cold dark matter as a major constituent of the Universe~\cite{1538-4357-648-2-L109,0067-0049-208-2-19}. Theoretical models propose the existence of weakly-interacting massive particles (WIMP) with masses in the range 1--15\,\gev\ as a possible explanation for dark matter~\cite{Cohen:2010kn}.  Statistically significant laboratory evidence for the detection of signals originating from scattering of WIMPs in light nuclear targets have been reported~\cite{PhysRevLett.111.251301,Bernabei:2013fk}.

The goal of the DAMIC experiment is to use the bulk silicon of a scientific-grade charge-coupled device (CCD) as the target for coherent WIMP-nucleus elastic scattering. The relatively low mass of the silicon nucleus, as well as the low read-out noise of the detector, make the CCD an ideal instrument for the identification of the nuclear recoils with keV-scale energies from WIMPs with masses $<$10\,\gev. Historically, CCDs had not been considered as a viable WIMP detector due to their relatively low mass ($\ll$1\,g per detector). Yet, recent advances in CCD technology, mostly due to the increase in the purity of the silicon, allow for the fabrication of  1--5\,g fully-depleted CCDs with exceptionally low levels of radioactive contamination. These instruments have been successfully characterized and deployed in astronomy experiments~\cite{Mclean:2012pka}, where their large area and thickness allow for the efficient detection of near-infrared light from astronomical objects.

The first WIMP search with a CCD as a WIMP target was performed at the MINOS near-site in Fermilab, yielding the best exclusion limits at the time for WIMPs with masses below 4\,\gev~\cite{Barreto2012264}. In November, 2012 the DAMIC collaboration deployed six 1\,g CCDs in the J-Drift hall of the SNOLAB laboratory in Canada. The vacuum vessel for this setup and the shielding were designed to house up to 1\,kg of silicon detectors. A subsequent deployment of two upgraded CCDs was performed in June, 2013 to address an unexpected source of uranium background in the CCD support. A further deployment of a setup with three new CCDs, expected to achieve radioactive background levels comparable to those necessary for a WIMP search to test the CDMS-Si signal~\cite{PhysRevLett.111.251301}, will take place in February, 2014. The full deployment of DAMIC100 at SNOLAB, consisting of eighteen 5.5\,g CCDs, is scheduled for Summer, 2014.

\section{Overview of a DAMIC CCD}
\label{sec:overview}

\begin{figure}[t!]
\begin{center}
\subfigure[A CCD pixel]{
\includegraphics[width=0.33\textwidth]{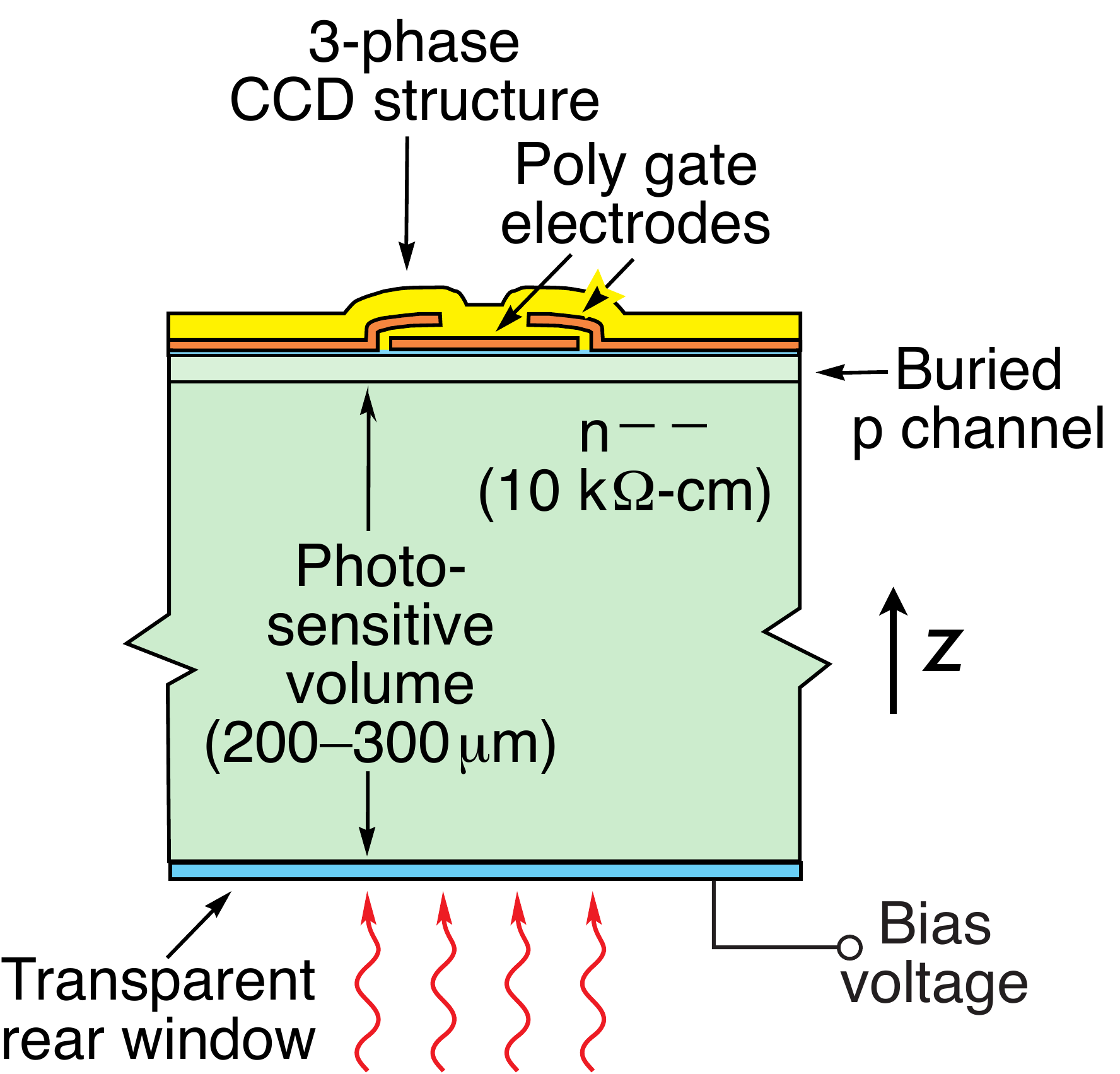}
\label{fig:pixel_sketch}
}
\hspace{4mm}
\subfigure[WIMP detection principle]{
\includegraphics[width=0.6\textwidth]{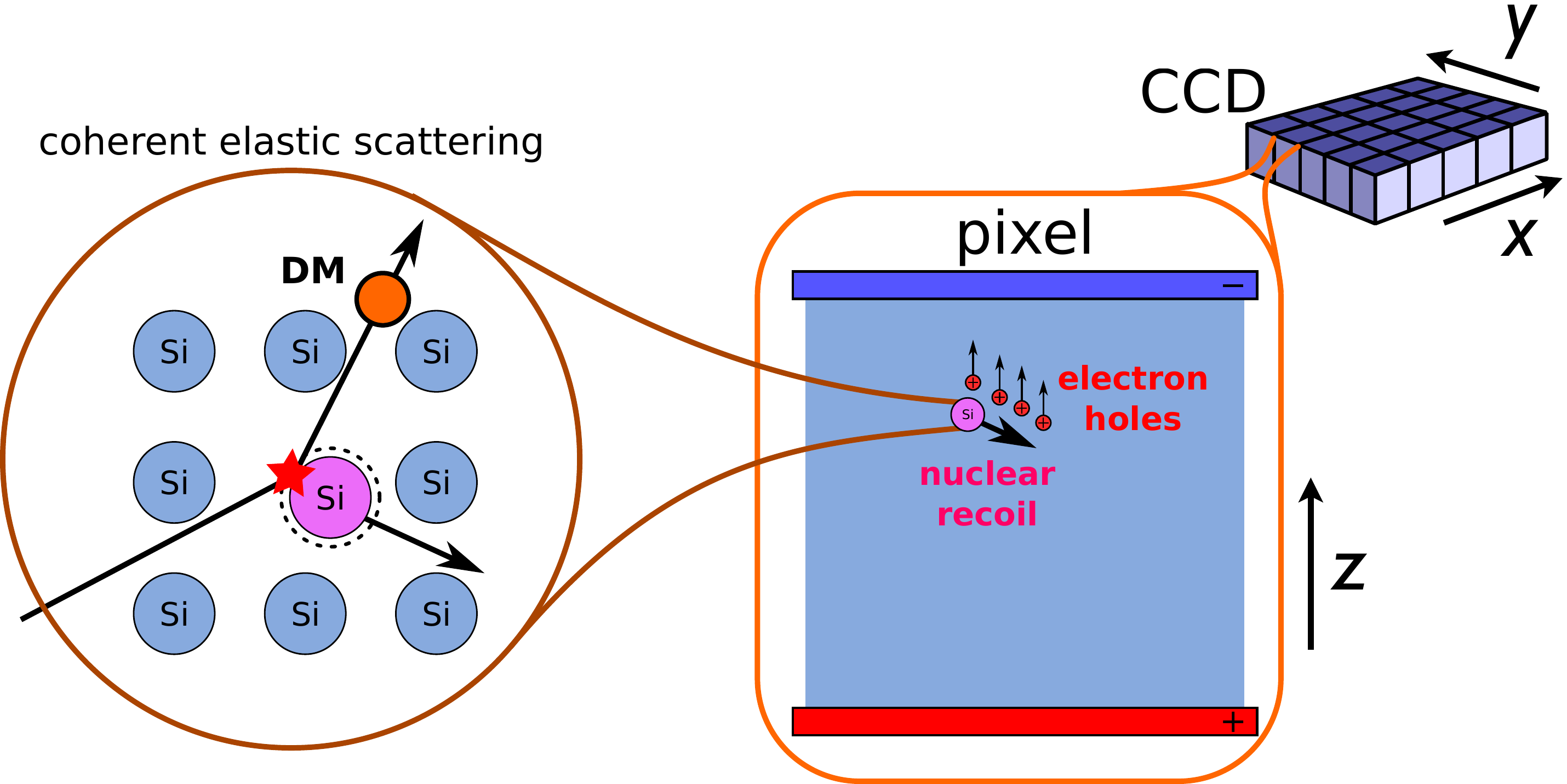}
\label{fig:ccd_sketch}
}
\end{center}
\caption{a)~Cross-sectional diagram of a 15\,$\mu$m\,$\times$\,15\,\um\ pixel in a fully depleted, back-illuminated charge-coupled device (CCD)~\cite{1185186}. The thickness of the gate structure and the backside ohmic contact are $\sim$0.1\um. As used for DECam, the transparent rear window includes an anti-reflective coating of induim-tin-oxide (ITO), which contains the undesirable radioactive isotope \indium. b)~Depiction of the WIMP detection principle, where the scattering of a DM particle with a Si nucleus leads to ionization being produced in the bulk silicon. The charge carriers are then drifted along the $z$ direction and collected at the CCD gates.
\label{fig:pixel_ccd_sketch}}

\begin{center}
\includegraphics[width=0.8\textwidth]{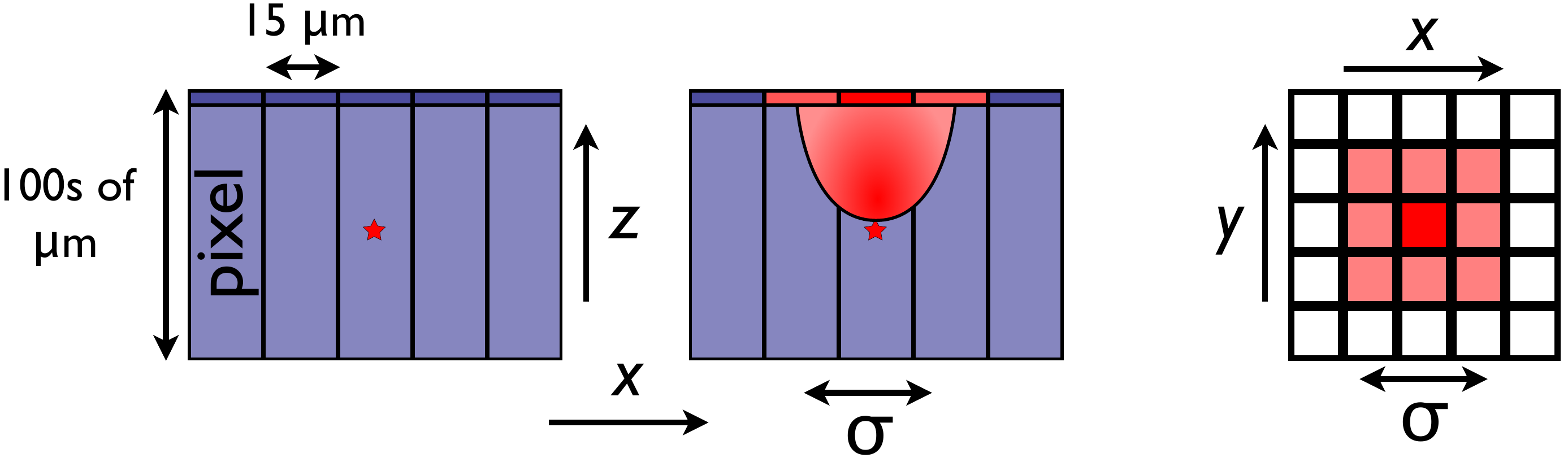}
\end{center}
\caption{Depiction of a point-like particle interaction within the CCD bulk. The charge is drifted along the $z$ axis and it diffuses as it travels toward the gates. This leads to a spatial distribution of the charge on the $x$-$y$ plane whose variance ($\sigma^2$) is proportional to the charge transit time. From this lateral spread it is possible to reconstruct the depth of the interaction. A characteristic value for the charge spread in a DECam CCD from the backside  (250\,\um\ deep) is $\sigma\sim7$\,$\mu$m.
\label{fig:diff_sketch}}
\end{figure}

The DAMIC CCDs were designed and fabricated at Lawrence Berkeley National Laboratory MicroSystems Lab for the Dark Energy Survey (DES) camera (DECam)~\cite{1185186}. They feature a three-phase polysilicon gate structure with a buried p-channel. The pixel size is 15\,$\mu$m\,$\times$\,15\,\um\ and the active region of the detector is high-resistivity (10--20 k$\Omega$\,cm) n-type silicon with hundreds of \um\ thickness. The high-resistivity of the silicon allows for a low donor density in the substrate ($\sim$\powero{11}\,cm\up{-3}), which leads to fully depleted operation at reasonably low values of the applied bias voltage ($\sim$20\,V). The CCDs are typically 8 or 16 Mpixels, with surface areas of tens of cm\up{2}. Fig.~\ref{fig:pixel_ccd_sketch} shows a cross-sectional diagram of the a CCD pixel, together with a sketch depicting the WIMP detection principle.

When operated at full depletion, ionization produced in the active region will be drifted along the direction of the electric field ($z$ axis). The holes (charge carriers) will be collected and held near the p-n junction, less than a \um\ below the gates. The electrons are drained from the backside. Due to the mobility of the charge carriers, the ionized charge will diffuse as it is drifted, with a spatial variance that is proportional to the carrier transit time. Charge produced by interactions closer to the back of the CCD will have longer transit times, leading to greater lateral diffusion. From the lateral spread of the charge recorded on the CCD $x$-$y$ plane, we can reconstruct the $z$ position of the charge deposit. This is depicted in Fig.~\ref{fig:diff_sketch}.

\section{Energy threshold}
\label{sec:e_threshold}

\begin{figure}[t!]
\begin{center}
\subfigure[Distribution of pixel values]{
\includegraphics[width=0.53\textwidth]{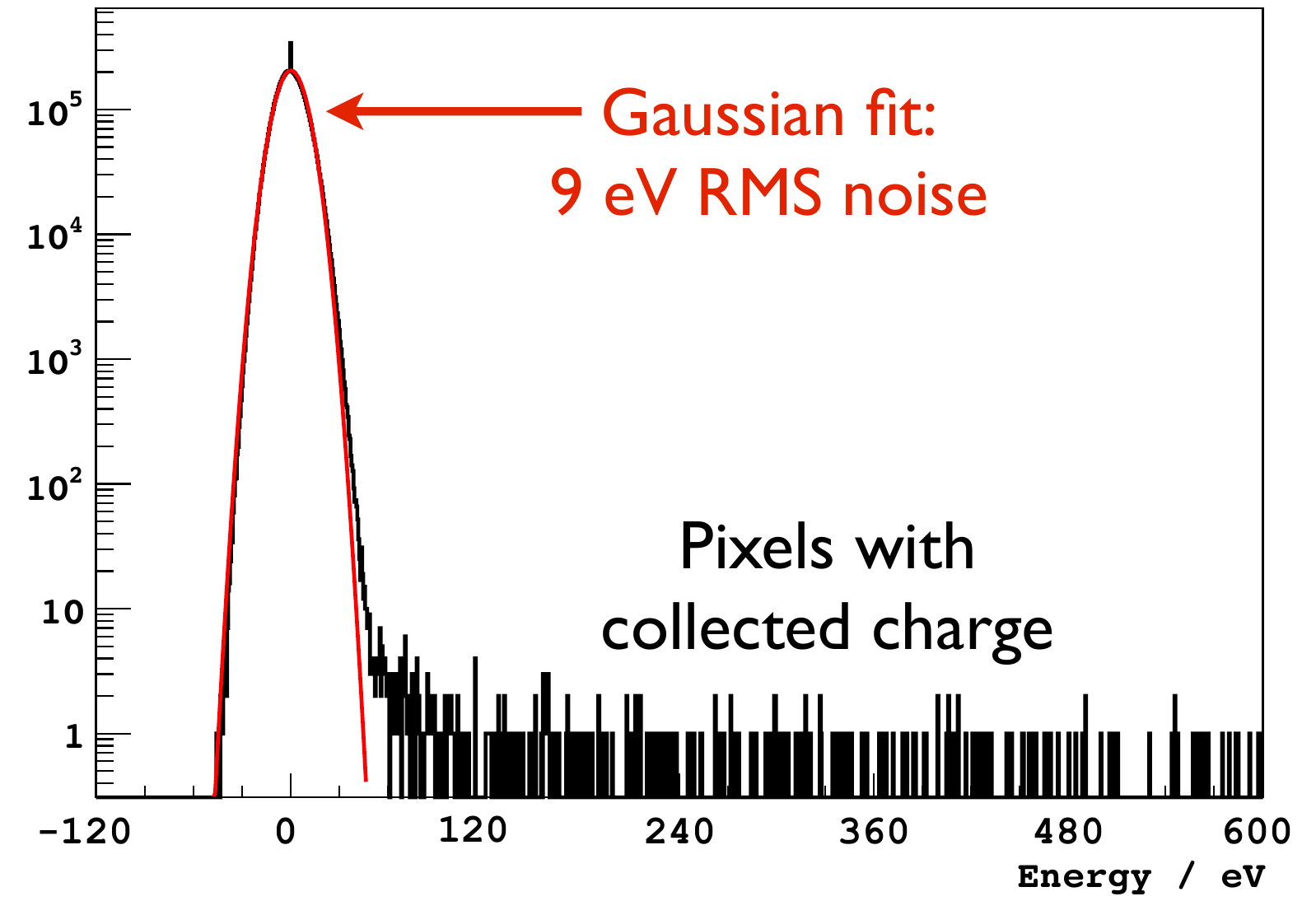}
\label{fig:pix_val_dist}
}
\hspace{2mm}
\subfigure[Segment of a DAMIC image]{
\includegraphics[width=0.38\textwidth]{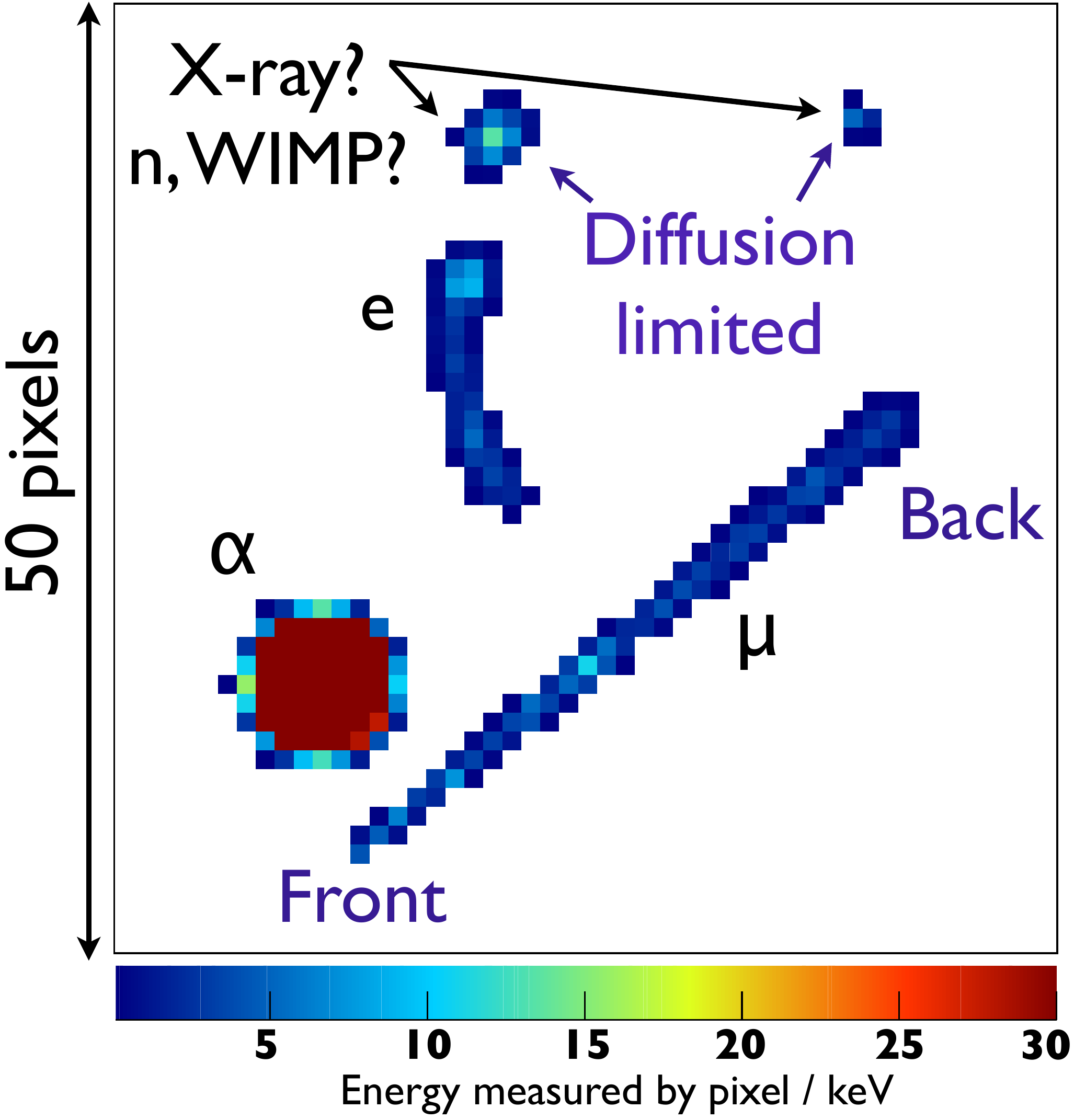}
\label{fig:image_eg}
}
\end{center}
\caption{a)~Histogram of all the pixel values in an image after the median pixel value over many images has been subtracted. The readout noise is the Gaussian distribution centered at zero, while the right-hand side tail corresponds to pixels where charge has been collected. The spike at zero is due to the image processing procedure, as every image is expected to contain some pixels whose value is the median over a set of images. b)~50$\times$50 pixel segment of a DAMIC image when exposed to a \cali\ source on the surface. Only pixels with deposited energy $>$0.1\,k\eve\ are colored. Clusters from different types of particles may be observed. Low energy electrons and nuclear recoils, whose physical track length is $<$15\,\um, produce ``diffusion limited" clusters, where the spatial extension of the cluster is dominated by charge diffusion (Fig.~\ref{fig:diff_sketch}). Higher energy electrons ($e$), from either Compton scattering or $\beta$ decay, lead to extended tracks. $\alpha$ particles in the bulk or from the back of the CCD produce large round structures due to the plasma effect~\cite{Estrada201190}. Cosmic muons ($\mu$) pierce through the CCD, leaving a straight track. The orientation of the track is immediately evident from its width, the end-point of the track that is on the back of the CCD is much wider than the end-point at the front due to charge diffusion.
\label{fig:image_pixel_val}}
\end{figure}

In DAMIC, the CCD is operated by applying the substrate bias across the active region and collecting the ionized charge over a few hours. Then readout is performed, where the charge is shifted row-by-row and the signal extracted through the serial register on one side of the CCD. As the capacitance of the output sense node of the CCD is very small, it is possible to measure only a few electrons of charge collected at the gates. The data stream is digitized and the pixel charge is measured. The read out rate is $\sim$1 Mpixel per minute. The RMS noise in each pixel measurement is $\sim$2.5 $e^-$ (Fig.~\ref{fig:pix_val_dist}). Considering that the average energy to create an electron-hole pair in Si is 3.62\,eV, this corresponds to 9~eV. From the measured pixel values an image is constructed, containing all ionization energy deposits within the CCD over the exposure time (Fig.~\ref{fig:image_eg}).

The number of dark electrons (i.e. those produced by thermal excitations in the Si substrate) collected in each pixel, which introduce Poissonian noise on its value, is proportional to the exposure time. For a 250\,\um\ DAMIC CCD running at 133\,K the dark current is typically $\sim$1\,$e^-$/pix/day. The exposure length of a DAMIC image is a few hours, therefore readout noise is the dominant source of noise.

As every image contains millions of pixels, to positively identify a pixel that has collected any charge, the condition that a pixel value is 5--6\,$\sigma$ above the noise level is required. This sets the nominal DAMIC threshold at $\sim$50\,\eve. The threshold for detection of an ionization event is variable, and depends on the number of pixels over which the charge is spread out, and consequently, on the depth of the interaction.

Due to the CCD operation, there is no time information on each of the detected energy deposits within a given exposure. Another noteworthy distinction is that, unlike other particle detectors, the number of events from instrumental noise in an event sample is not proportional to the exposure length but to the number of times that the CCD has been read out.

\section{Energy response}
\label{sec:e_response}

The charge loss in a CCD is dominated by charge transfer inefficiency (CTI) as the CCD is readout. This value is generally $\sim$\powert{-6}{1} for a DECam CCD~\cite{1185186}. Thus, $<$1\% of the charge is lost for the furthest-most pixel in an 8 Mpixel CCD. The well capacity of a CCD pixel is generally $>$\powero{5} electrons. Therefore, a pixel may collect up to a charge equivalent to $\sim$360\,k\eve. In practice, the limitation for the largest energy that can be measured in a pixel is set by the dynamic range of the digitizer. This value is 30\,k\eve\ under the standard settings of the DAMIC CCD readout. Due to charge diffusion and the relative low \dedx\ for electrons, it is unlikely for an ionizing electron to deposit $>$30\,k\eve\, in one pixel, leading to practically no limitation on the measurement of the energy deposited by an electron in the CCD. Nuclear recoils are more likely to reach this maximum value due to their higher \dedx, but considering their lower ionization efficiency~\cite{PhysRevA.45.2104}, the ionization energy of recoils up to 100\,k\evr\ can be measured precisely.

\begin{figure}[t!]
\begin{center}
\subfigure[Spectrum from a \ironfive\ source]{
\includegraphics[width=0.59\textwidth]{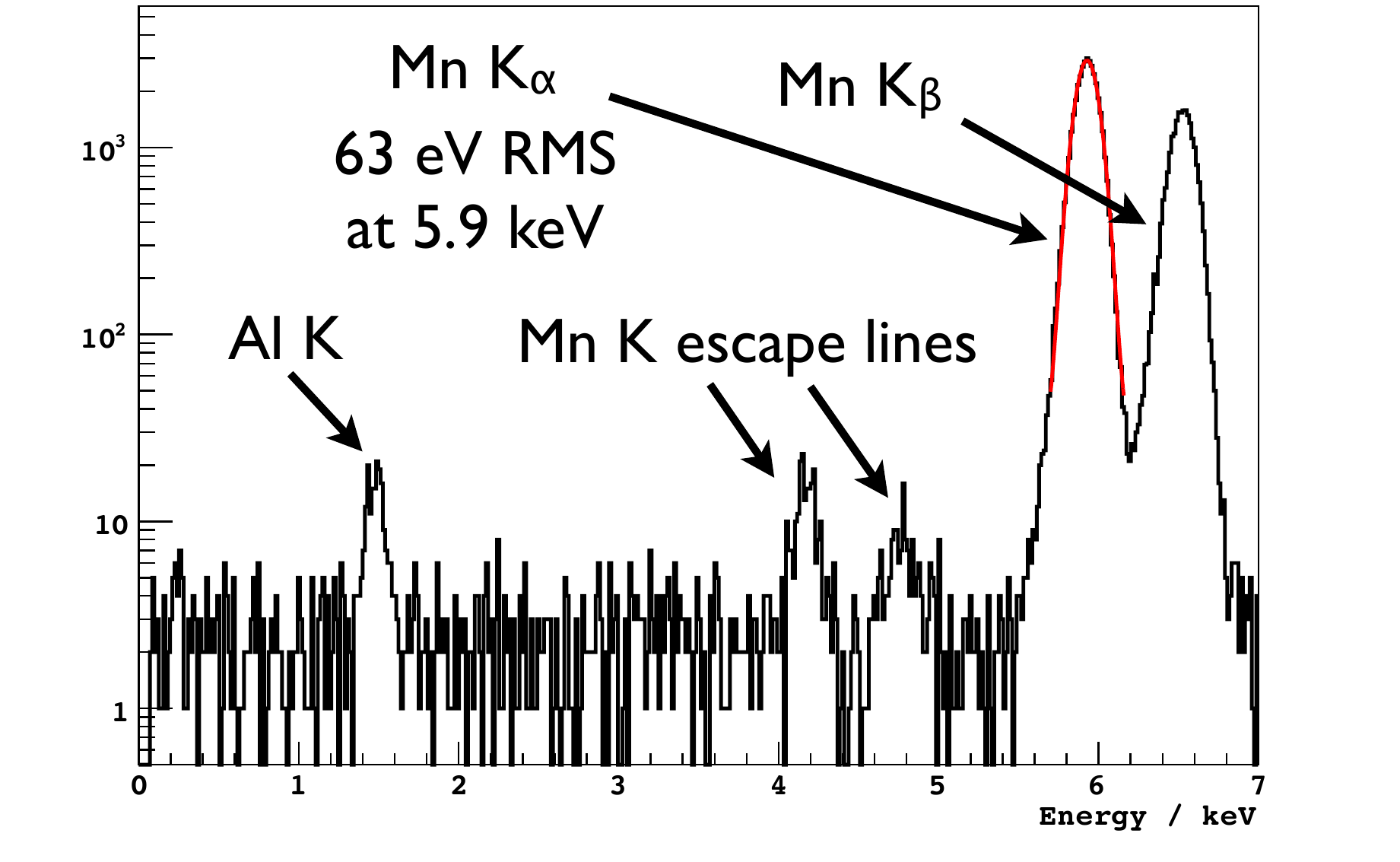}
\label{fig:fe55_fit}
}
\subfigure[Observation of a escape event]{
\includegraphics[width=0.33\textwidth]{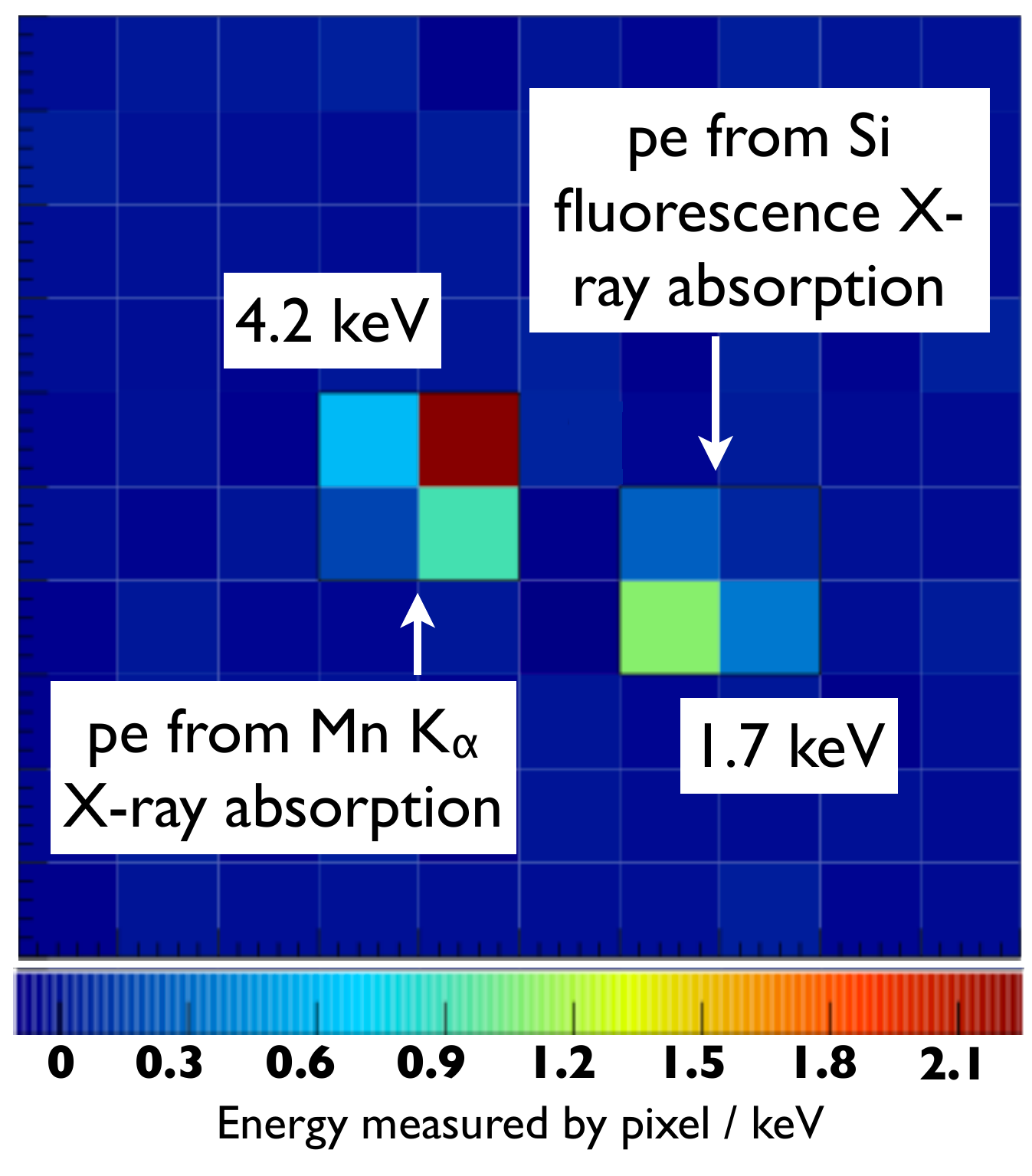}
\label{fig:escape_event}
}
\end{center}
\caption{a)~Spectrum obtained when the CCD is illuminated from the back with a \ironfive\ source. The two main X-rays from the source are the K$_\alpha$ and K$_\beta$ from the daughter Mn nucleus. The source holder is made of aluminum, which is the origin of the fluorescence Al K X-ray observed. The main Mn K$_\alpha$ and K$_\beta$ peaks are formed by X-rays that deposit their full energy in the CCD, while the Mn escape lines are due to partial energy deposits, where the subsequent Si fluorescence X-ray (1.7\,keV) escapes the CCD. The absorption length of this X-ray is 14\,\um. b)~Rarely, it occurs that the fluorescence X-ray travels far enough within the CCD (a few attenuation lengths) so that its energy deposit is pixels away from the first X-ray interaction, leading to two distinguishable clusters, as shown here. 
\label{fig:fe55}}
\end{figure}

\begin{figure}[t!]
\begin{center}
\subfigure[Linearity of the ionization energy scale]{
\includegraphics[width=0.48\textwidth]{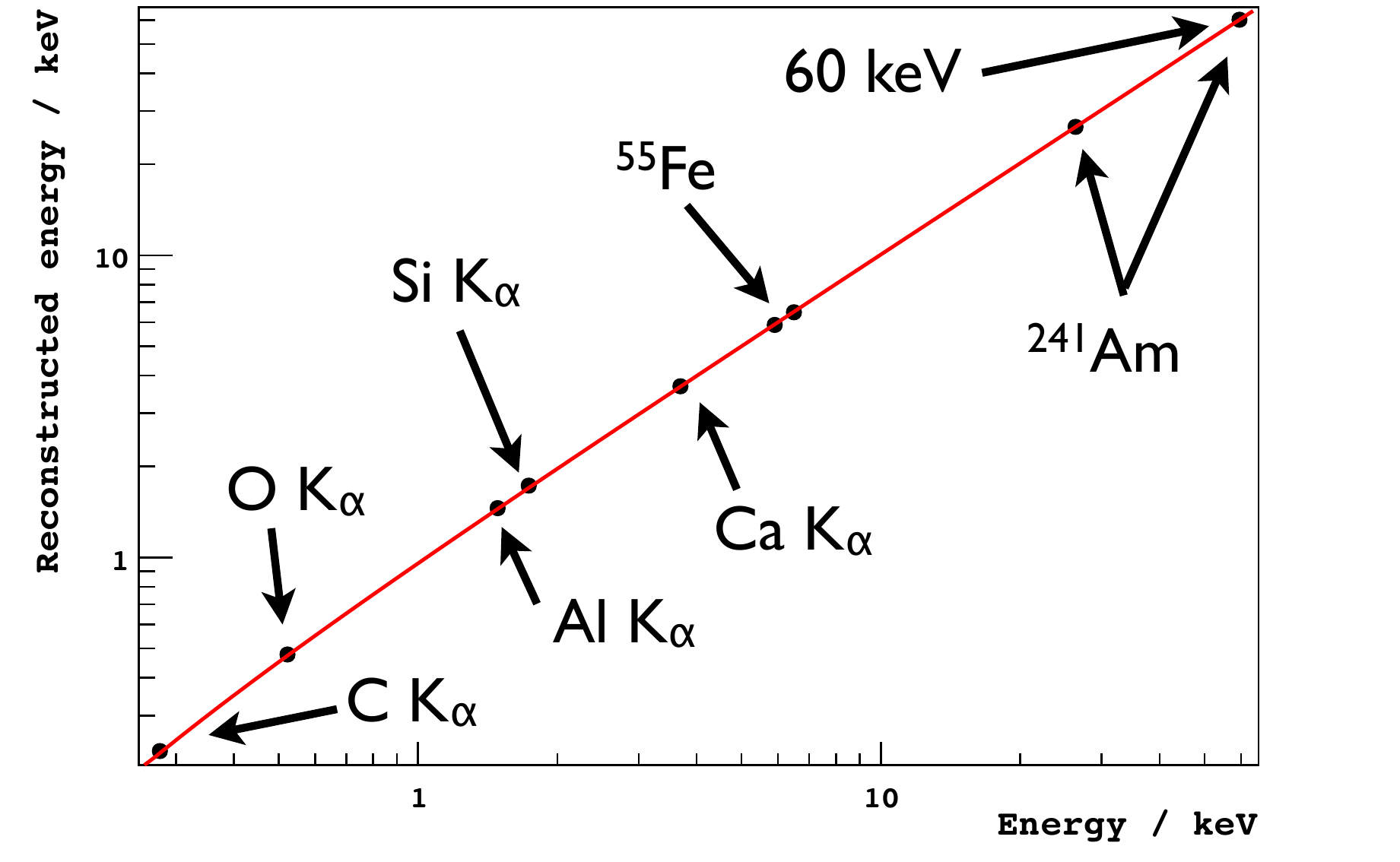}
\label{fig:elinearity}
}
\subfigure[Energy resolution]{
\includegraphics[width=0.48\textwidth]{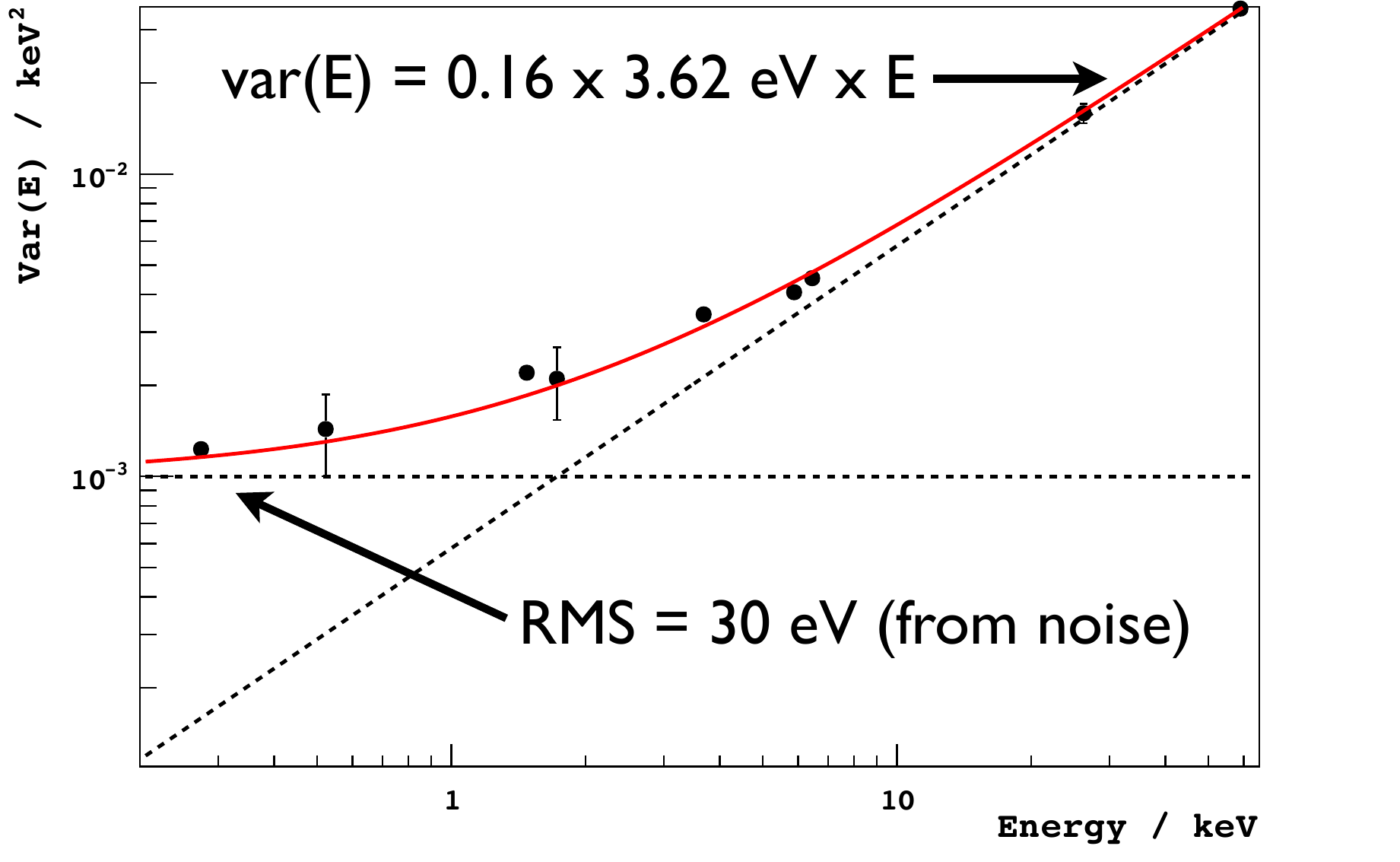}
\label{fig:eres}
}
\end{center}
\caption{a)~Reconstructed energy of an X-ray line compared to its true energy. The labeled K$_\alpha$ markers are fluorescence lines from elements in the Kapton target and other materials in the CCD setup. The \ironfive\ and \amer\ markers are X-rays emitted by the radioactive sources. Linearity in the measurement of ionization energy is demonstrated from 0.3\,k\eve\ to 60\,k\eve. 
b)~Variance of the X-ray lines as a function of energy. The effective Fano factor is 0.16, larger than the accepted value in Si of 0.1, but typical for a CCD~\cite{janesick2001scientific}. As the illumination is from the backside, the charge is spread over many pixels. Thus, the readout noise added over many pixels leads to a limiting resolution of 30\,\eve.
\label{fig:eresponse}}
\end{figure}

The energy spectrum measured by a DAMIC CCD from exposure to a \ironfive\ source is shown in Fig.~\ref{fig:fe55_fit}. Calibrations with fluorescence X-rays from a Kapton target exposed to the \ironfive\ source or $\alpha$s from \amer\ were also performed. Fig.~\ref{fig:elinearity} demonstrates the linearity of the CCD in the measurement of ionization energy produced in the active region, while Fig.~\ref{fig:eres} quantifies the energy resolution of the CCD. Detection of the  0.28\,k\eve\ carbon K$_\alpha$ X-ray (attenuation length of 0.14\,\um) is possible because the inert surface of the CCD is only $\sim$0.1\um\,thick. For these calibrations the CCD was illuminated from the back, which, due to charge diffusion, leads to a larger dynamic range and worse energy resolution than if the CCD had been illuminated from the front.

As the CCD is fully depleted, there are no field-free regions that may lead to partial charge collection. 
The observed line shapes and spectra from X-ray sources placed in the front and in the back of the CCD are consistent with the absence of any population of events with a significant loss of collected charge.

It is well known that the ionization efficiency of nuclear recoils is significantly different than that of electrons. Previous measurements have been done down to energies of 3--4\,k\evr~\cite{PhysRevA.45.2104,PhysRevD.42.3211}, yielding results in agreement with Linhard theory~\cite{ziegler1985stopping}. From this, DAMIC's nominal 50\,\eve\ threshold corresponds to $\sim$0.5\,k\evr. Given the significant uncertainty in the extrapolation, and the importance of precise nuclear recoil scale calibration for dark matter searches, we are planning a series of experiments to measure this value down to the threshold (Section~\ref{sec:calibration}).

\section{Position reconstruction}
\label{sec:pos_recon}

Due to the pixelated nature of the CCD, the best estimate for the $x$ and $y$ coordinates of a point-like interaction may be readily obtained from the charge-weighted mean of the $x$ and $y$ coordinates of pixels with collected charge. In the worst case, where the entire charge is collected in one pixel, the resolution in the values of the $x$ and $y$ coordinates is 4.3\,\um. For clusters where the charge is distributed over many pixels the resolution gets better, to $\lesssim$1\,\um. The observation of escape events within the CCD (Fig.~\ref{fig:escape_event}) is a demonstration of its capability to resolve energy deposits 10s of \um\ apart. 

\begin{figure}[t!]
\begin{center}
\subfigure[Spatial $\sigma$ distribution for Mn K$_\alpha$]{
\includegraphics[width=0.48\textwidth]{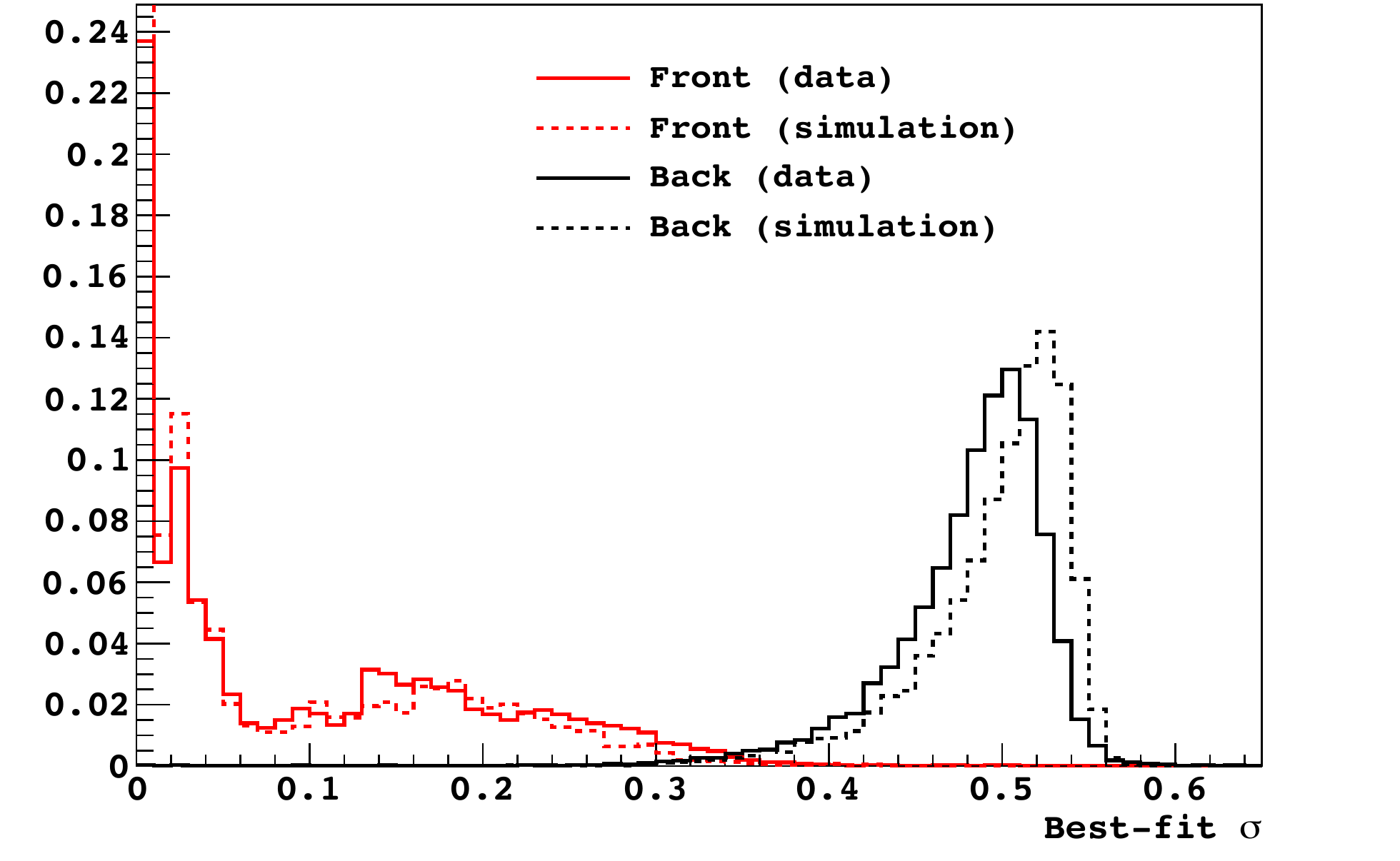}
\label{fig:fe55_sigma}
}
\subfigure[Spatial $\sigma$ vs. interaction depth ($z$) for Mn K$_\alpha$]{
\includegraphics[width=0.48\textwidth]{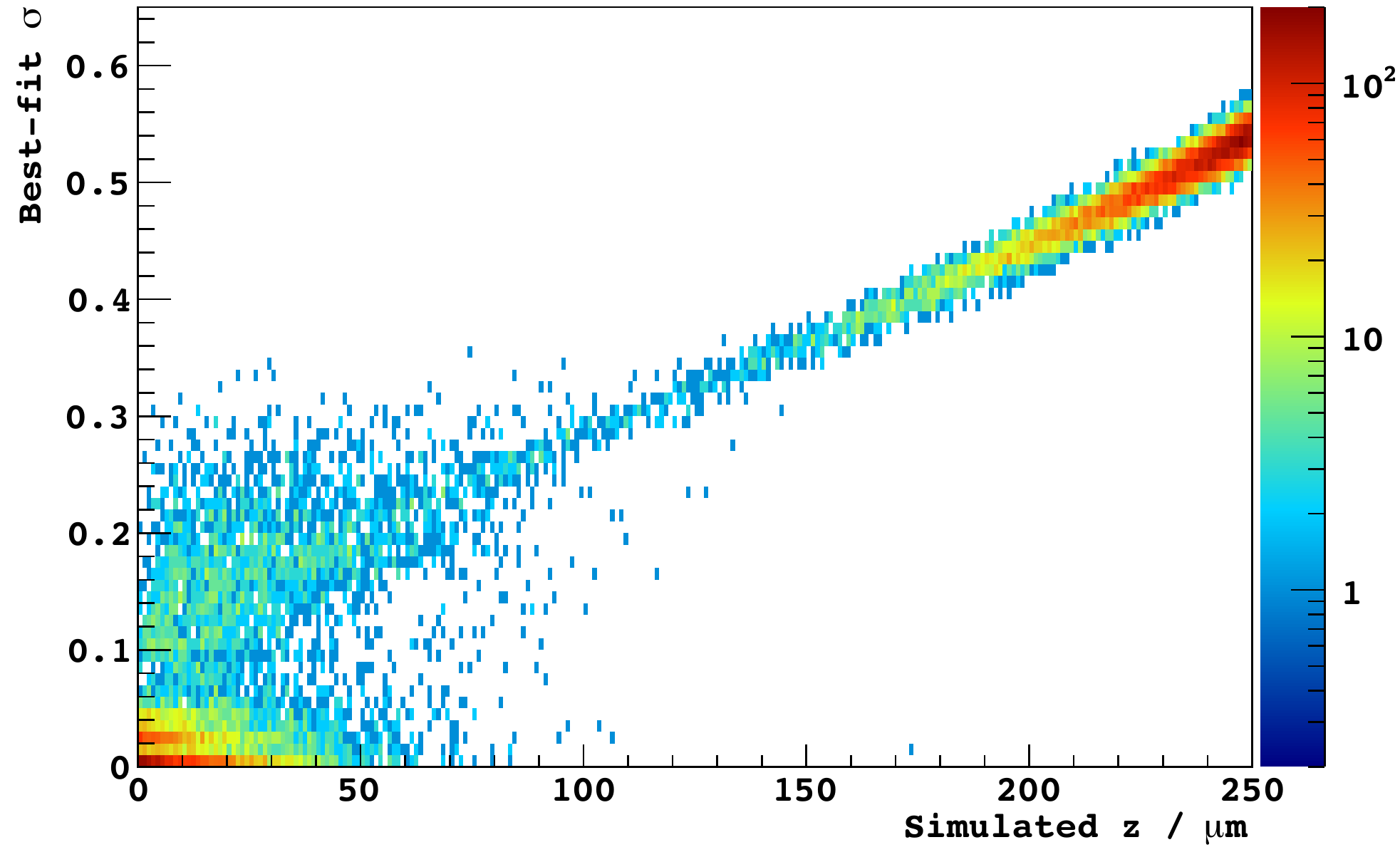}
\label{fig:fe55_simz}
}
\end{center}
\caption{a)~Best-fit $\sigma$ (in pixels) distribution for Mn K$_\alpha$ (5.9\,keV) X-rays for the cases where a \ironfive\ source is placed in the front (red) and in the back (black) of a DAMIC CCD. The solid lines represent calibration data acquired at Fermilab, while the dashed lines are the results from the simulation. The simulation was tuned to a different CCD deployed at SNOLAB, therefore the mismatch between the distributions is partly due to performance variations between CCDs. The overall shape of the distributions is well reproduced by the simulation. b)~Relationship between the best-fit $\sigma$ and the interaction depth. $\sigma$ was obtained by performing the fit on simulated clusters, where the simulated $\sigma$ was computed from the interaction depth ($z$). The interaction depth was obtained from a MCNP particle physics simulation of a \ironfive\ source on a Si target with the CCD geometry. The expected positive correlation between $\sigma$ and $z$ is recovered. Clusters on the back of the CCD may be easily identified as $\sigma$ is obtained reliably. For cases where the charge is spread over very few pixels, $\sigma$ cannot be precisely measured, leading to worsening determination of $\sigma$ at small values of $z$.
\label{fig:fe55_recon}}
\end{figure}

\begin{figure}[t!]
\begin{center}
\subfigure[Low energy X-rays from the back]{
\includegraphics[width=0.48\textwidth]{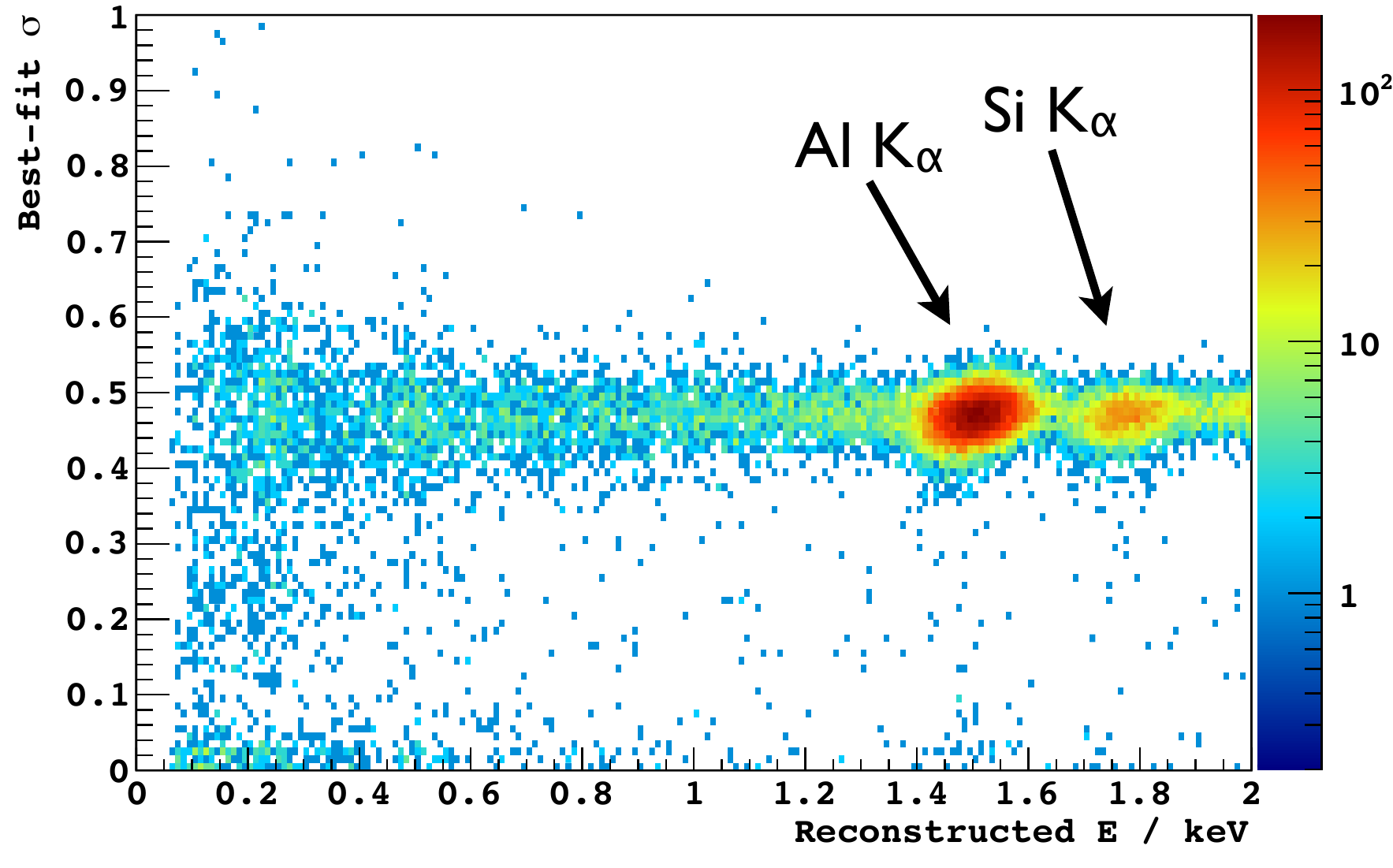}
\label{fig:lowe_sigma_e}
}
\subfigure[\cali\ source (uniform)]{
\includegraphics[width=0.48\textwidth]{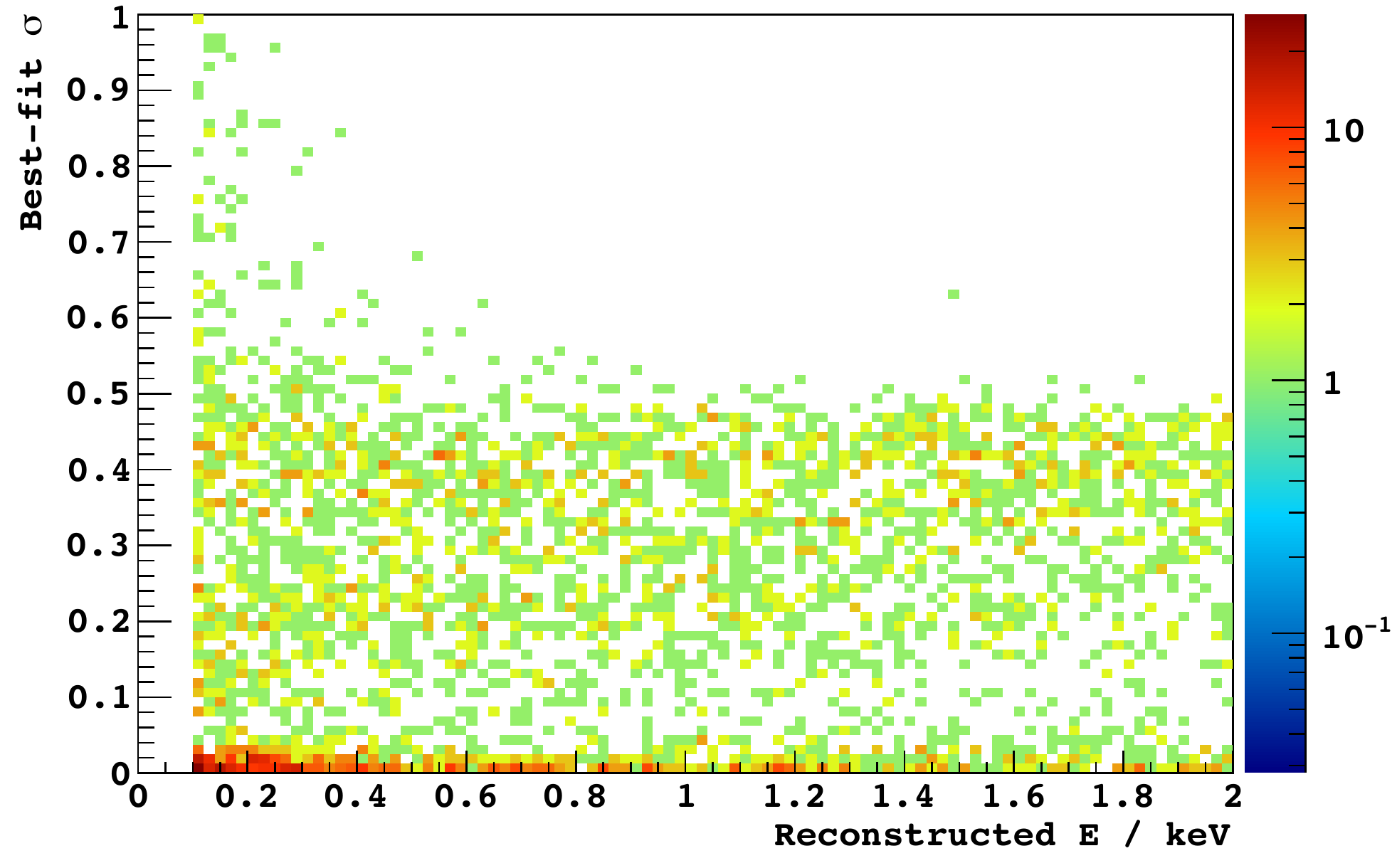}
\label{fig:cf_sigma_e}
}
\end{center}
\caption{a)~Best-fit $\sigma$ (in pixels) against electron-equivalent energy for clusters detected when the CCD is illuminated from the back with fluorescence X-rays from a Kapton target exposed to a \ironfive\ source. The absorption length for X-rays in this energy range is $<$10\,\um. b)~Best-fit $\sigma$ (in pixels) against electron-equivalent energy for clusters detected when the CCD is exposed to fast neutrons from a \cali\ source. Due to the neutrons' large interaction length ($\gg$CCD thickness) these events are distributed uniformly in the CCD bulk. The difference between the $\sigma$ distribution of surface events on the back of the CCD and of uniformly distributed events is evident down to the threshold.
\label{fig:sigma_e}}
\end{figure}

As discussed in Section~\ref{sec:overview}, the spatial spread ($\sigma$) in a diffusion limited cluster may be used to reconstruct the depth ($z$ coordinate) within the active region where the energy deposit took place. To estimate the best value for $\sigma$, we perform a likelihood fit to all the pixels within a four pixel radius of the pixel in the cluster with the largest value. We assume a Gaussian distribution in two dimensions, with $\sigma$ and the mean values of $x$ and $y$ as free parameters. We consider the readout noise on every pixel when doing the fit. To test the performance of this procedure, we have constructed a corresponding simulation, where events are generated on the true CCD readout noise pattern and following the hypothesized Gaussian spatial distribution of the collected charge. The simulated value for $\sigma$ is based on the depth of the interaction (obtained from an independent MCNP particle physics simulation) assuming a simple model for the electric field in a fully depleted CCD~\cite{1185186}. The field magnitude has been tuned so that the maximum observed diffusion matches that from a particular CCD deployed in SNOLAB. Fig.~\ref{fig:fe55_recon} presents the $\sigma$ reconstruction results for the Mn K$_\alpha$ X-ray. Fig.~\ref{fig:sigma_e} shows the best-fit $\sigma$ distributions as a function of energy for calibration data with low energy X-rays impinging on the back of the CCD, and for events from a \cali\ source, which are expected to be uniformly distributed in the bulk.

\section{DAMIC setup at SNOLAB}
\label{sec:snolab}

\begin{figure}[t!]
\begin{center}
\includegraphics[width=\textwidth]{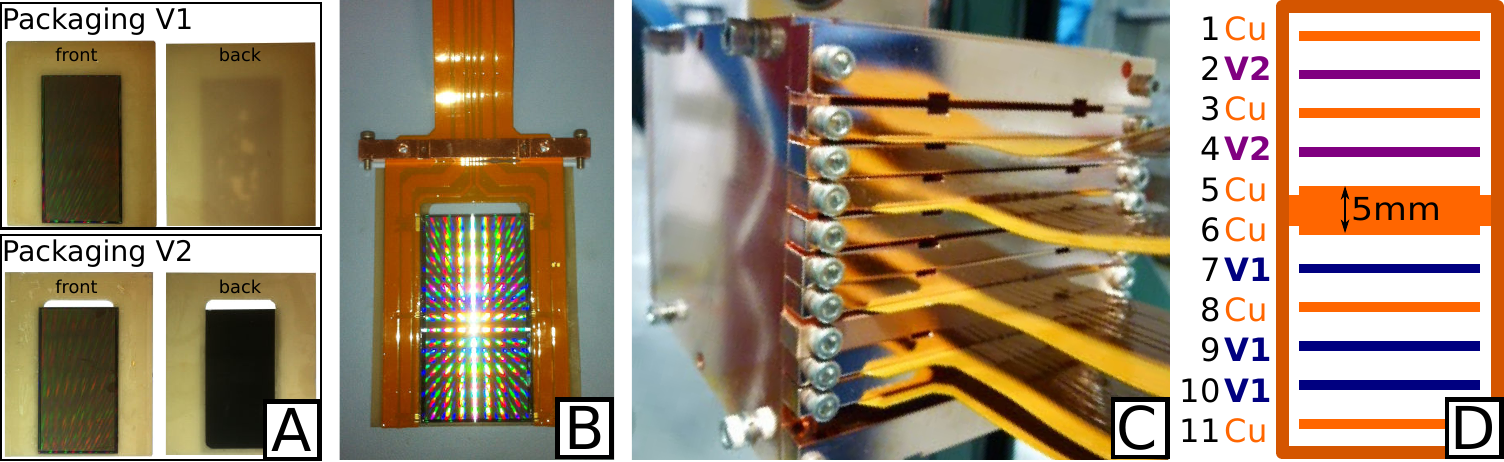}
\end{center}
\caption{A)~Version V1~(top) and V2~(bottom) of the CCD package developed for the SNOLAB tests. The CCDs are 6\,cm$\times$3\,cm$\times$250\,\um, 8 Mpixel CCDs of the same batch as those used for DECam. The CCDs are epoxied on an aluminum nitride (AlN) support piece. In the later version the AlN mass is reduced from 17\,g (V1) to 11\,g (V2) by removing most of the substrate material in contact with the active area of the CCD, leaving only an AlN frame.
  B)~Complete package with long Kapton cable, which brings the signal outside the shield to the electronics.
  C)~Stack of five CCD detectors inside the copper box ready for installation into the vacuum vessel for operation at SNOLAB.
  D)~Ordering of the CCD detectors inside the copper box for the second deployment in June, 2013. The two CCDs using package V2 are enclosed between copper slabs. The CCD in slot 10 is facing up, it has AlN support below and above it (from the CCD in slot 9). The first deployment of DAMIC in November, 2012 consisted of six CCDs with package V1 and no copper slabs in between. 
\label{fig:snolab_setup}}
\end{figure}

\begin{figure}[t!]
\begin{center}
\includegraphics[width=0.8\textwidth]{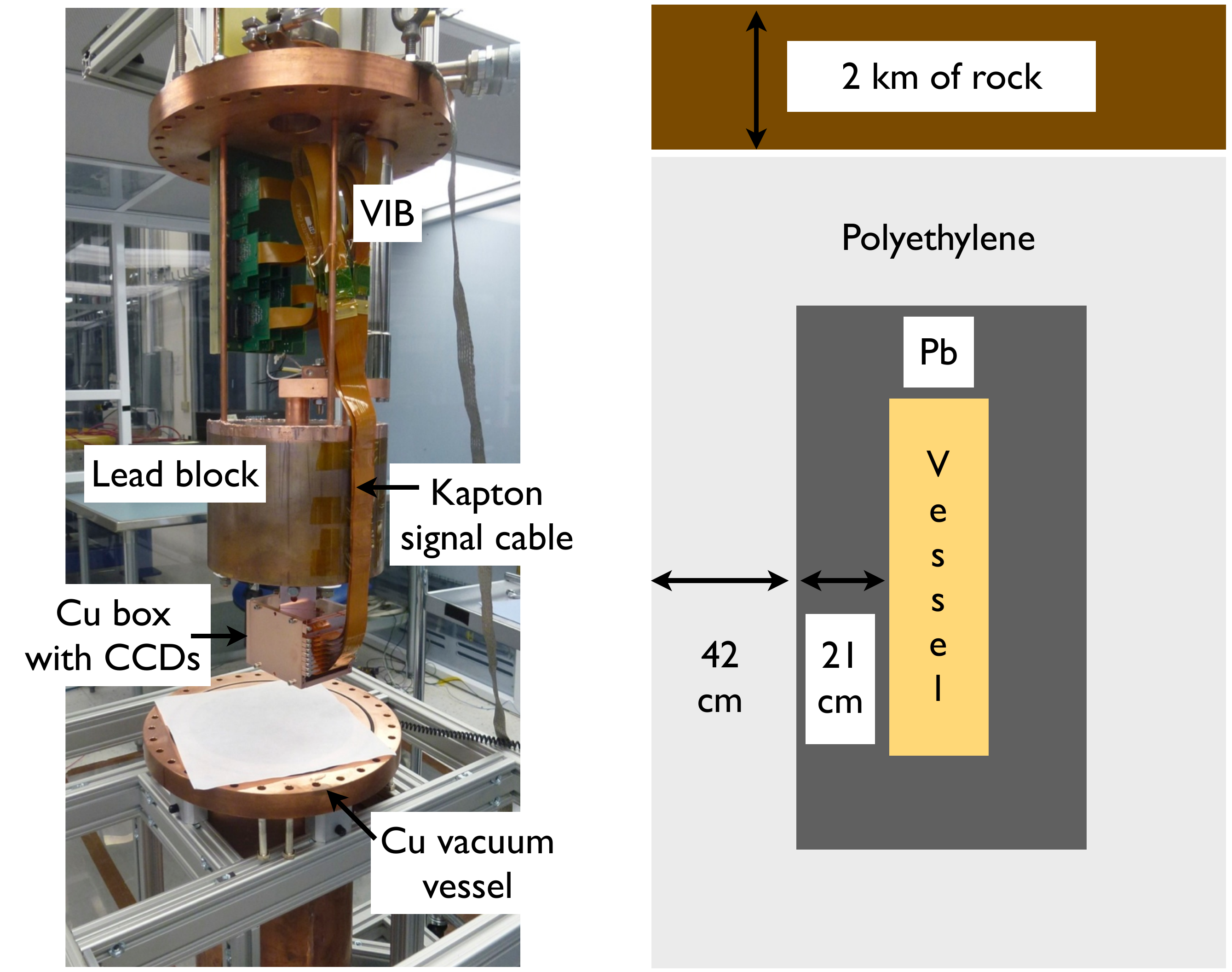}
\end{center}
\caption{The copper box holding the CCDs (Fig.~\ref{fig:snolab_setup}) is placed within a copper vessel, where they are kept in a \powero{-7}\,torr vacuum. This vessel is large enough to accommodate up to 1\,kg of CCDs. Within the vessel, above the CCD box, there is a copper-laminated block of lead, 21\,cm high, which serves as a shield from radiation from the vacuum-interface board (VIB) immediately above it. There is a hole in the shield traversed by a copper cold finger, which keeps the copper box at the CCD operating temperature of 133\,K. Thin, Kapton signal cables fit through the small spacing between the internal Pb shield and the inner side of the vacuum vessel to bring the signals from the CCDs up to the VIB. The vessel is placed within a 21\,cm Pb shield (\pbten\ rate $=$ 58$\pm$18\,Bq/kg), to stop external $\gamma$-rays. Beyond the Pb shield there are 42\,cm of high-density polyethylene to attenuate external neutrons. On one side of the shielding there is a hole a few inches in diameter (not shown), through which signal cables from the VIB board go to the outside electronics crate. The helium lines from the outside compressor to the cold head also go through this hole.
\label{fig:snolab_shield}}
\end{figure}

DAMIC was deployed in SNOLAB in November, 2012. An upgrade to address the observed uranium background in the AlN support piece took place in June, 2013. Fig.~\ref{fig:snolab_setup} shows the arrangement of the DAMIC inner detector in these deployments. Fig.~\ref{fig:snolab_shield} depicts the shielding in which the inner detector is housed.

\begin{figure}[t!]
\begin{center}
\subfigure[SNOLAB spectrum, 1--30\,k\eve]{
\includegraphics[width=0.48\textwidth]{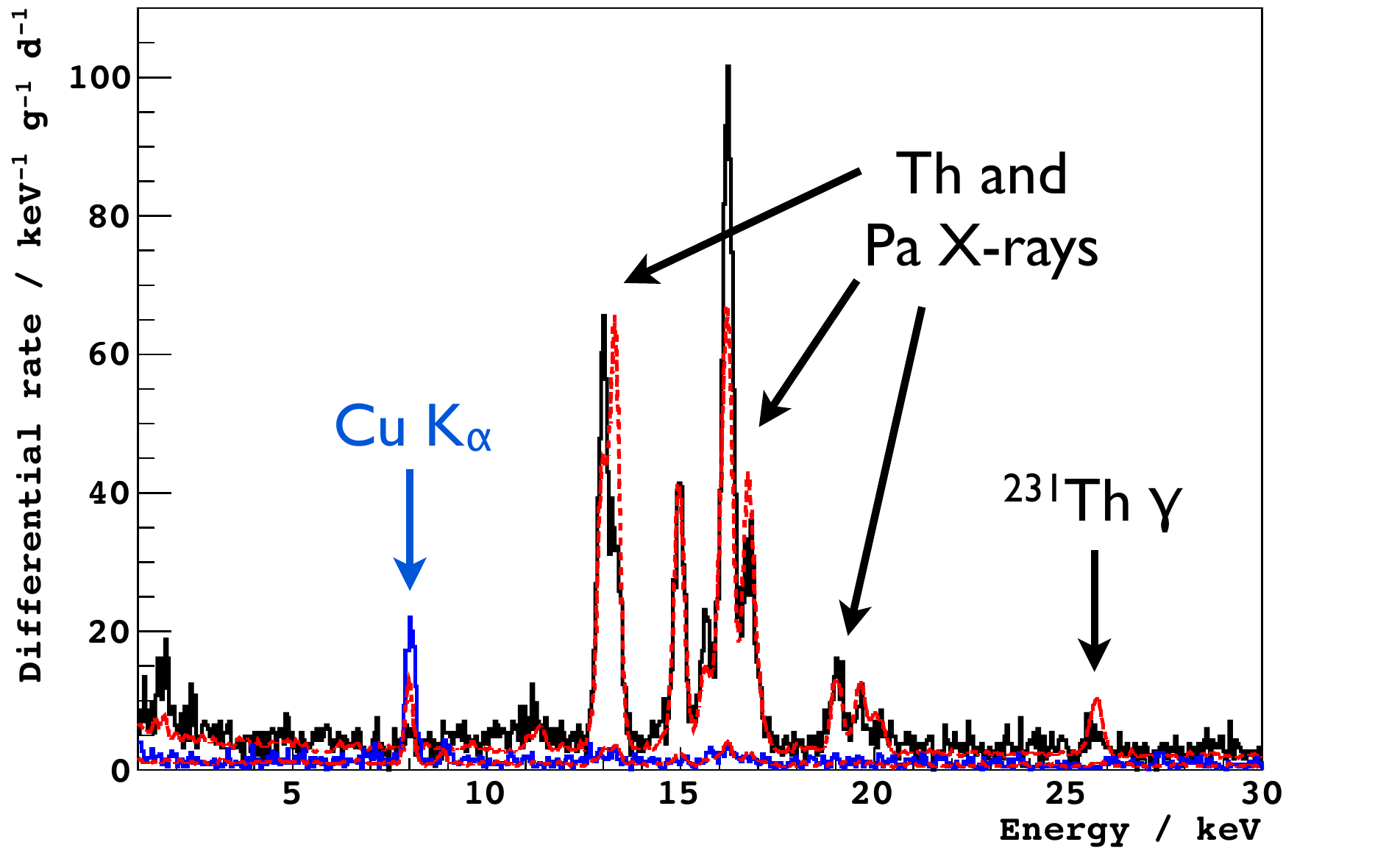}
\label{fig:spec_5-30}
}
\subfigure[SNOLAB spectrum, 30--500\,k\eve]{
\includegraphics[width=0.48\textwidth]{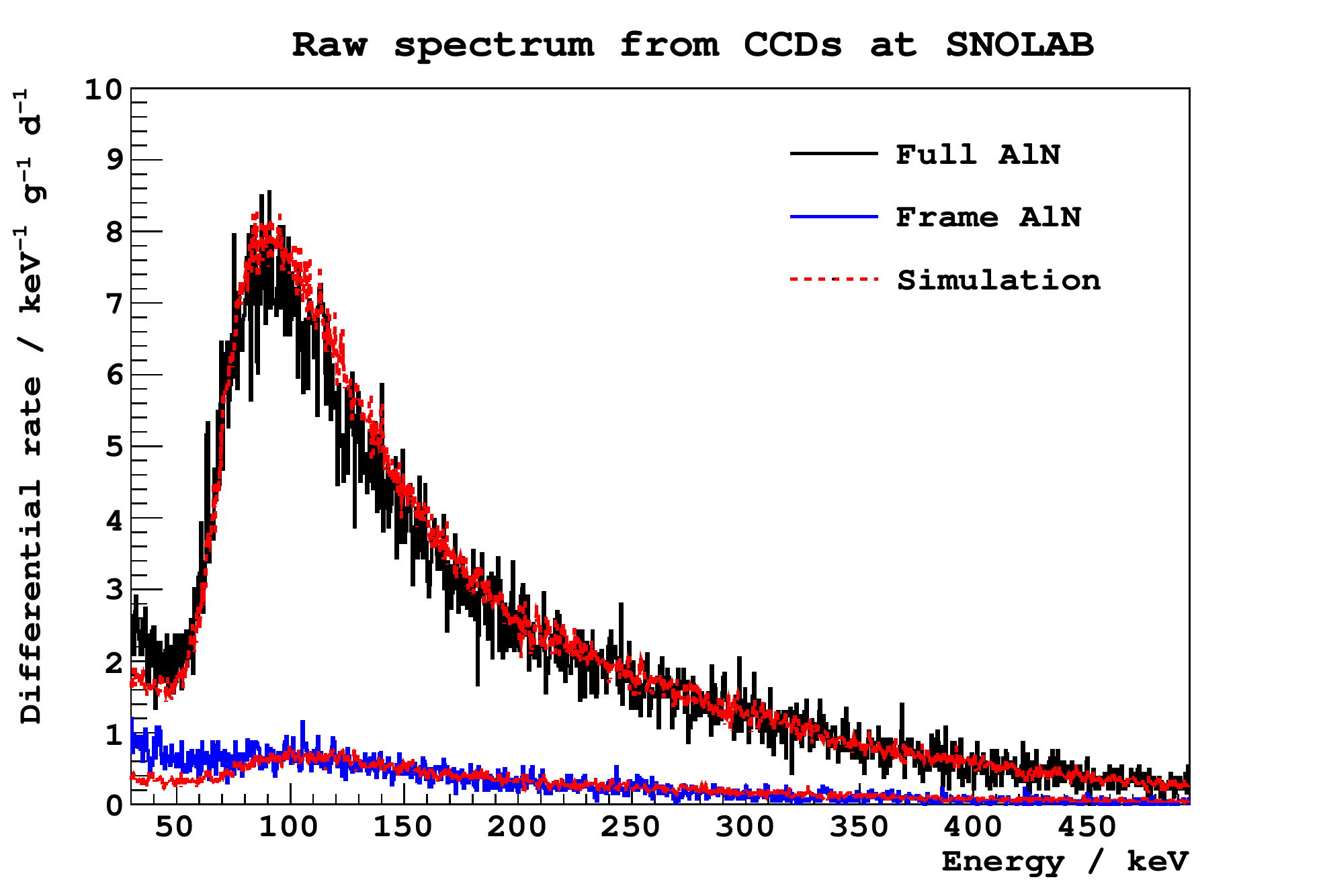}
\label{fig:spec_30-500}
}
\end{center}
\caption{Differential rate spectrum for all clusters in the first deployment of DAMIC (black) and in the CCDs with the frame support in the second deployment (blue). The red line is the spectrum obtained by simulating with MCNP the energy deposits in the CCD from radioactive decay in the AlN. The simulation has been fit to the black histogram, with the absolute intensities of the decay chain segments \ura--\urafour\ and \urafive--$^{231}$Th, and the 14.9\,keV fluorescence line from yttrium (present in the AlN at the \% level) as free parameters.
The minimum ionizing (MI) bump is a generic spectral feature due to MeV-scale electrons that do not deposit their full energy in the CCD, which is well reproduced by the simulation. The simulation is also run with the frame AlN geometry in packaging V2. The simulated spectrum in the CCD originating from the contaminations in the AlN determined from the fit to the CCD with packaging V1 is overlaid on the blue line. There is a general agreement on the decrease in the level of the background. The Cu K fluorescence line produced by radiation from the AlN that impinges on the copper slabs surrounding the CCDs with V2 packaging also appears in the simulation.
\label{fig:snolab_spec}}
\end{figure}

In the original design for DAMIC, the AlN was deemed adequate for the detector's radioactive background requirements due to the estimated 10\,ppb of \ura\ from $\gamma$-ray screening of \rad\ daughters, and the assumption of secular equilibrium~\cite{Groom:2002fk}. Unfortunately, when DAMIC at SNOLAB was switched on, a background at the level of \powert{3}{5}\,\dru\ was observed at low energies ($<$10\,k\eve). Characteristic X-rays from uranium decay were immediately visible (Fig.~\ref{fig:snolab_spec}), and subsequent $\gamma$-ray screening measurements performed at SNOLAB directly measured the \ura\ content to be 330\,ppb (4.1\,Bq/kg), strongly out of secular equilibrium. The modification of the AlN geometry into a frame in the subsequent detector deployment lead to a significantly decreased background (Fig.~\ref{fig:snolab_spec}).

\section{Analysis of SNOLAB data}
\label{sec:analysis}

We present the data collected at SNOLAB by two CCDs (1.1\,g each) with V2 packaging (Fig.~\ref{fig:snolab_setup}) between June 6 and September 2, 2013 (52\,days of live time). The first 40 days of live time were acquired in 5\,ks exposures, while the latter 12 days in 10\,ks exposures. The analysis focuses on the energy region $<$10\,k\eve, as for a WIMP search. We begin by identifying ``seed" pixels whose collected charge is at least four times the RMS value of the noise in an image, corresponding to $\sim$40\,\eve\ of ionization energy. We proceed to find the cluster of pixels around  the seed whose values are at least twice the RMS noise. Each cluster is then considered as a candidate for a particle interaction. Variables are computed for each cluster and the following data selection cuts are performed to select physical events in the CCDs:

\begin{enumerate}[(i)]
\item A minimum energy threshold cut is applied as a function of the number of pixels in the cluster. This is to exclude the presence of clusters from correlated noise, where a few adjacent pixels may be systematically higher than the median of the noise.
\item A goodness-of-fit selection is performed from the result of the fitting procedure to extract the cluster spread ($\sigma$) (Section~\ref{sec:pos_recon}). This is a cut on the maximum likelihood of the cluster to be described by a symmetric, two-dimensional Gaussian distribution of the pixel values, as expected for physical, diffusion-limited events. 
\end{enumerate}

The clustering efficiency together with the acceptance of these cuts as a function of energy was evaluated from both the \cali\ data and from simulated events on top of ``blanks" (i.e. short exposures acquired at SNOLAB with very few, if any, physical events but with the true readout noise patterns). The first 40\,days of data had a higher noise level, with 50\% acceptance at 0.3\,k\eve, while the last 12\,days had 50\% acceptance at 0.1\,k\eve\ and full acceptance for energies $>$0.15\,k\eve.

\begin{figure}[t!]
\begin{center}
\subfigure[Best-fit $\sigma$ for events from CCDs in V2 packaging]{
\includegraphics[width=0.48\textwidth]{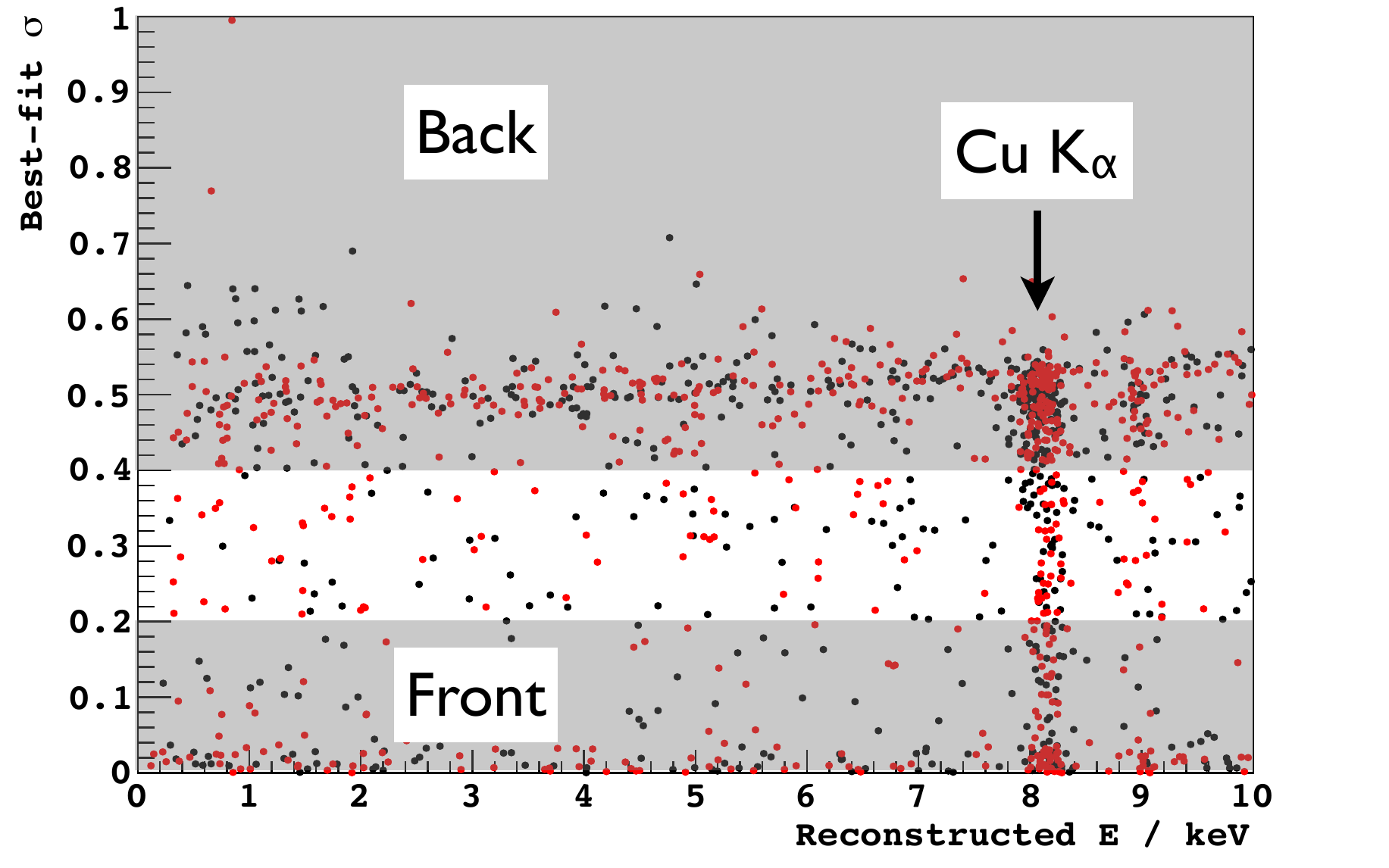}
\label{fig:snolab_sigma_e}
}
\subfigure[Spectrum after selection of bulk events]{
\includegraphics[width=0.48\textwidth]{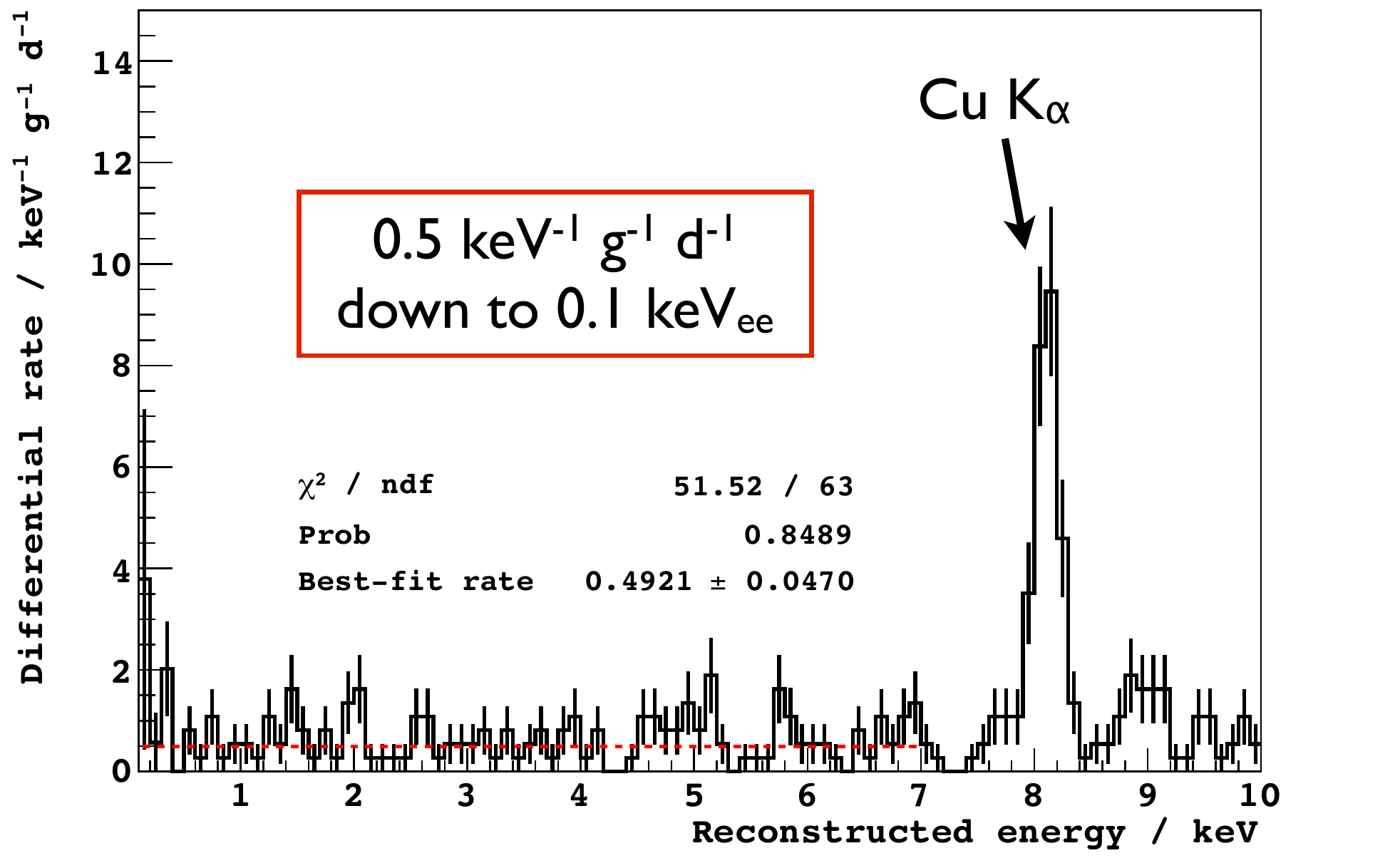}
\label{fig:snolab_e_final}
}
\end{center}
\caption{a)~Best-fit $\sigma$ (in pixels) against electron-equivalent energy for events in CCDs with V2 packaging and energies $<$10\,k\eve\ (full spectrum is blue histogram in Fig.~\ref{fig:snolab_spec}). Events in the gray region of the plot are expected to be mostly from surface events. The Cu K events are incident from both the front and the back of the CCD but, due to their 43\,\um\ absorption length, a significant fraction of them penetrate into the bulk. b)~Spectrum of events in the bulk (white) band in Fig.~\ref{fig:snolab_sigma_e}. We have corrected for the acceptance of bulk events after the $\sigma$ selection cut, obtained from the \cali\ calibration data (Fig.~\ref{fig:cf_sigma_e}). A fit to a flat spectrum has been performed in the region 0.1--7\,k\eve\ to obtain the limiting background in the bulk.
\label{fig:snolab_final}}
\end{figure}

Finally, a fiducial cut is performed, by selecting events with 0.2\,pix$<$$\sigma$$<$0.4\,pix, as depicted in Fig.~\ref{fig:snolab_sigma_e}. The acceptance of this cut for bulk events is 35\%, while the leakage from surface events is 20\% near threshold and improves to 10\% at 0.7\,k\eve. The bulk events are distributed uniformly on the $x$-$y$ plane and the residual bulk background is measured to be  \powert{2}{5}\,\dru\ (Fig.~\ref{fig:snolab_e_final}). Even though some contribution from uranium decays in the AlN may still be present, this residual background is consistent with the expectation from Compton scattering of bremsstrahlung photons produced in the lead shield by the decay of the daughter of \pbten, \biten\ (58$\pm$18 Bq/kg).

\section{Radioactive contamination in the CCD}
\label{sec:backgrounds}

So far there is no evidence for any radioactive contamination in the CCDs.

Clusters produced by $\alpha$ decays were observed throughout the physics run of the DAMIC CCDs. There is a large variation in their rate between CCDs, and it is highest for the CCD directly below the full AlN support (slot 10 in Fig.~\ref{fig:snolab_setup}(D)), likely due to the $\alpha$s from the \ura\ decays immediately above. $\alpha$s from the AlN below are stopped by the 100\,\um\ of epoxy between the CCD and its support. The lower rate of $\alpha$s observed in the other CCDs are probably due to \pbten\ contamination on the CCD surface and on the copper surfaces above and below the CCD. Due to the limited dynamic range of the current data set, it is not possible to measure the energy of these $\alpha$s. Later in 2013 we plan an extended run with a decrease in the charge integration window, leading to a substantial increase in the digitizer range that will allow us to perform $\alpha$ spectroscopy. A few month run should allow us to set a limit on the \ura\ and \tho\ contamination in the CCD bulk $<$100\,ppt. For the \ura\ limit we would consider the number of $\alpha$ decays with energies in the range 4.6\,MeV -- 4.9\,MeV, corresponding to the decays of $^{234}$U, $^{230}$Th, and $^{226}$Ra. For \tho\ we would search for a $\sim$19\,MeV deposit from the spatial pile up of the fast $\alpha$ decay sequence $^{224}$Ra $\rightarrow$ $^{220}$Rn $\rightarrow$ $^{216}$Po. 

The imaging capabilities of the CCD should also allow us to constrain the highly uncertain \pbten\ and \sitwo\ contamination in the silicon, by relying on the fact that these decays and the decays of their respective daughters, \biten\ and \ptwo, should take place in the same spatial position. We will be performing a search for two electron tracks from $\beta^-$ decay that start in the same pixel on different images. Given that the half-lives of \biten\ and \ptwo\ are much longer than the exposure times (days vs. hours), and that the high spatial resolution of the CCD will greatly suppress accidentals, it should be possible to place decay rate limits  $<$mBq/kg, even with modest background levels. Likewise, this technique offers a very powerful and unique tool for background suppression in case that either of these two contaminants turns out to be a significant, limiting contribution to the count rate in the bulk.

\section{Near-future and DAMIC100}
\label{sec:upgrade}

To address the current radioactive background issues of DAMIC, a deployment of new detectors is scheduled for February, 2014. An upgrade of the lead shield, with the addition of an inner layer of ancient lead (\pbten\ rate $<$0.02 Bq/kg) will follow in May, 2014. In this iteration three new 8 Mpixel detectors (7.2\,g total) will be installed in the copper box. The CCDs will be epoxied to a high-purity silicon support piece. The Kapton flex cable will remain wire-bonded as in Fig.~\ref{fig:snolab_setup}(B). Two of these CCDs will be 500\,\um\ thick astronomical CCDs with an ITO anti-reflective coating on the backside, while the third will be 650\,\um\ thick without ITO, very similar to the CCDs to be used for DAMIC100. For the astronomical CCDs, the dominant background is expected to come from \indium\ decays 
in the ITO at a rate of 6 events/(CCD$\cdot$day). The 650\,\um\ CCD is expected to observe a count rate of $\sim$\dru, close to the goal for DAMIC100.

\begin{figure}[t!]
\begin{center}
\subfigure[Spectrum after 1\,y of DAMIC100]{
\includegraphics[width=0.48\textwidth]{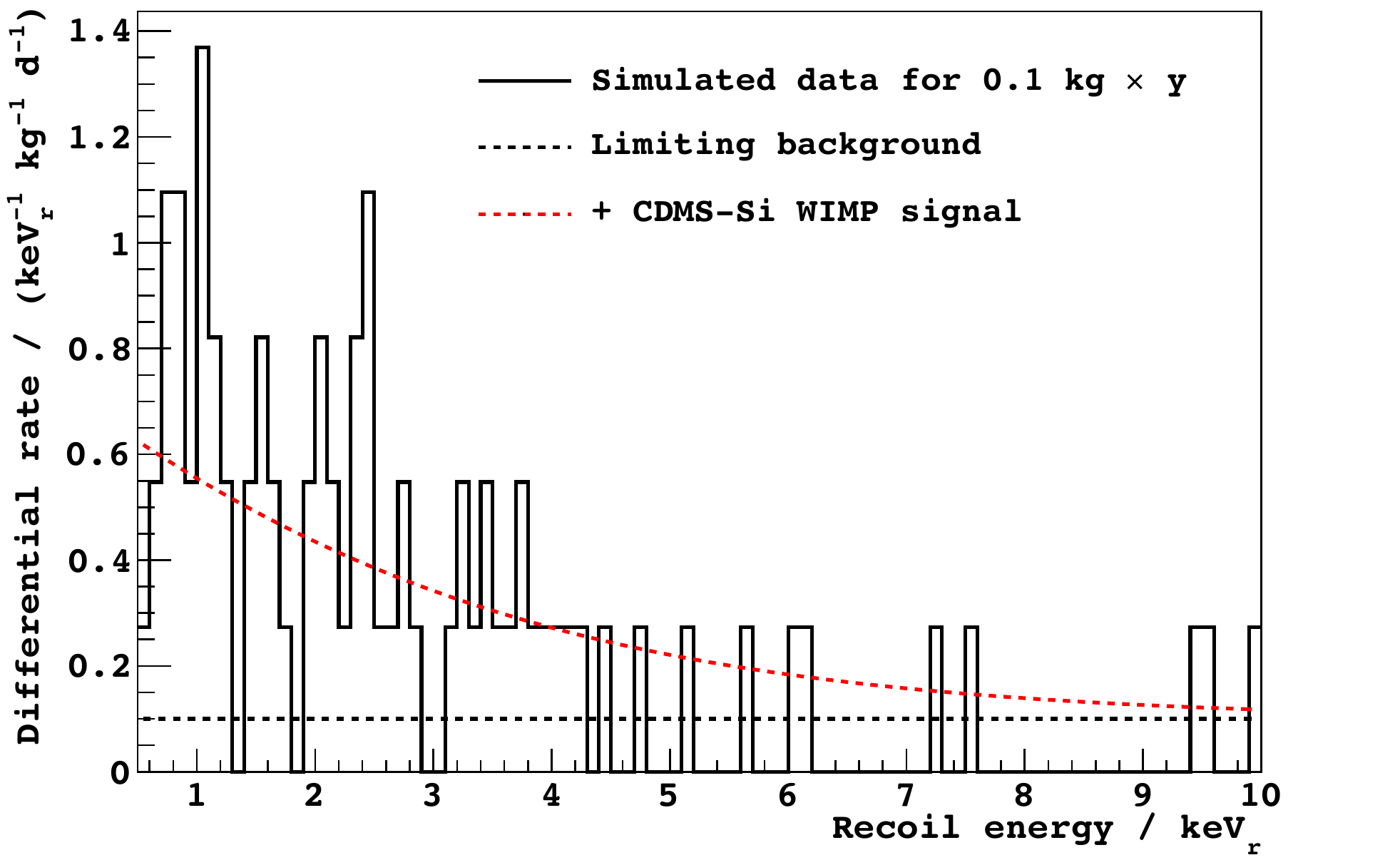}
\label{fig:damic100_spec}
}
\subfigure[Exclusion plot after 1\,y of DAMIC100]{
\includegraphics[width=0.46\textwidth]{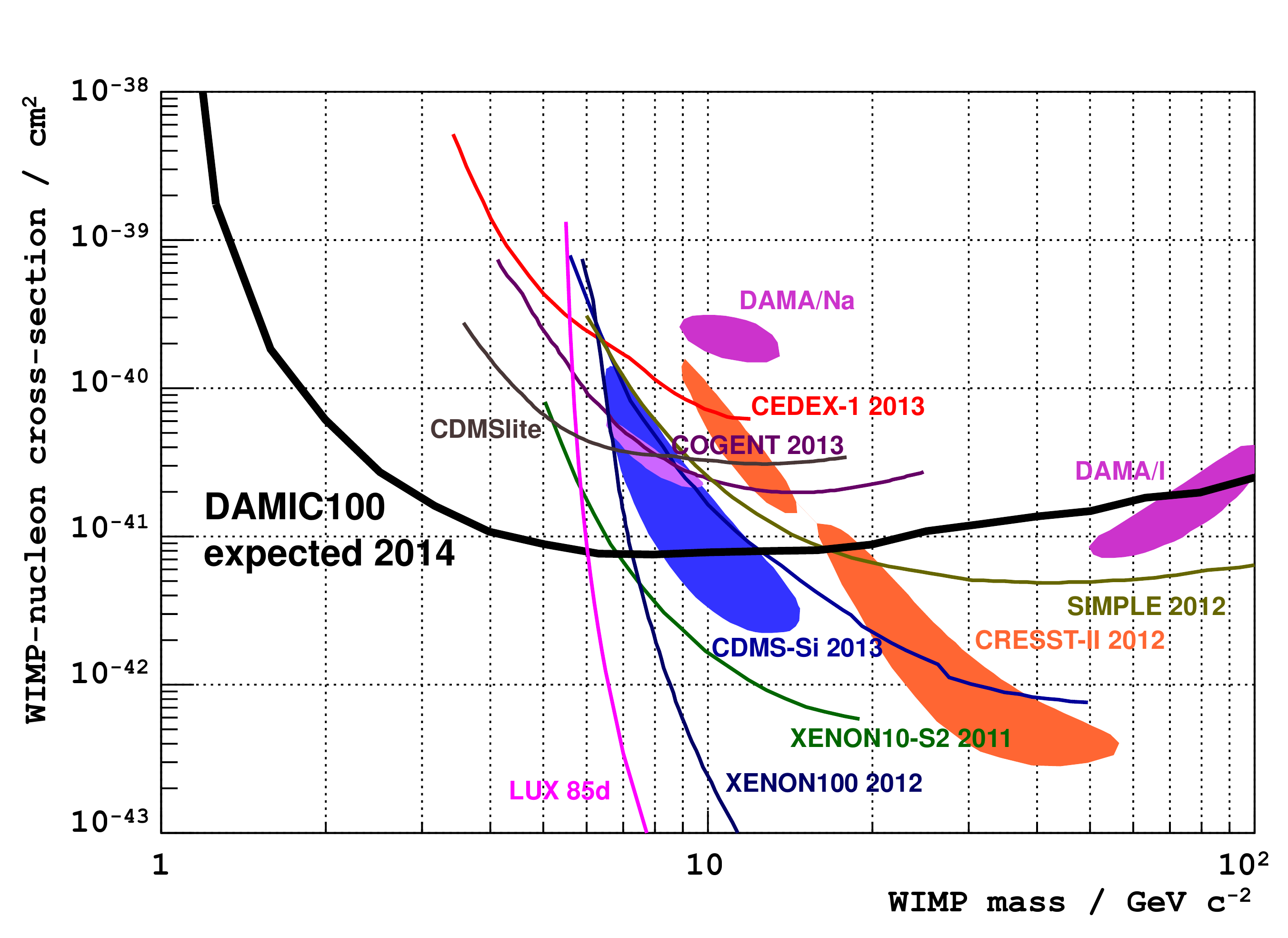}
\label{fig:exclusion}
}
\end{center}
\caption{a)~Simulated spectrum of DAMIC100, considering a WIMP with the mass and interaction cross-section of the best-fit to the CDMS-Si signal ($M_\chi$$=$8.6\,\gev\ and \swn$=$\cst{41}{1.9})~\cite{PhysRevLett.111.251301}, standard halo parameters ($\rho_\chi$$=$0.3\,\gev/cm$^{3}$, $v_0$$=$220\,m/s, $v_E$$=$232\,m/s, $v_{esc}$$=$544\,m/s) and a 0.1\,kg$\cdot$y exposure. For this illustration, the ionization efficiency of nuclear recoils is assumed to be 0.2 and energy independent. Thus, the expected limiting background of 0.5\,\dru\ corresponds to 0.1\,events/(keV$_{r}$$\cdot$kg$\cdot$d). The exponential increase toward low energies, starting below 5\,k\evr\ ($\sim$1\,k\eve), is evident. b)~Under these assumptions we present a 90\% exclusion plot for spin-independent interactions by performing a $\chi^2$ test on simulated spectra with the flat background spectrum and the simulated WIMP signal for different values of $M_\chi$ and \swn. DAMIC100 will place the best limits on spin-independent WIMP-nucleon elastic scattering for $M_\chi$$<$6\,\gev.
\label{fig:damic100_reach}}
\end{figure}

DAMIC100 will consist of eighteen, 650\,\um\ thick, 16 Mpixel LBNL CCDs with a total mass of 100\,g in our current vacuum vessel and shielding in SNOLAB. These detectors have been purchased and will start being delivered to Fermilab in May, 2014. They will be installed on high-purity silicon supports, immediately surrounded by OFHC copper. The radioactive background from the CCD packaging is expected to be $\ll$1\,\dru\ and the count rate should be dominated by Compton scattering from external $\gamma$-rays, at a predicted rate of 0.5\,\dru. Operation of the new detector should start in Summer, 2014. Fig.~\ref{fig:damic100_reach} shows the expected reach of DAMIC100 and the magnitude of the possible signal from the light WIMP scenario suggested by CDMS-Si~\cite{PhysRevLett.111.251301}.

\section{Calibration program}
\label{sec:calibration}

Extensive calibration has been performed on DAMIC CCDs at Fermilab. We have exposed the CCDs to X-ray, $\gamma$-ray and fast neutron sources to characterize their response to ionization energy (Section~\ref{sec:e_response}) and to the diffusion of events from the surface and in the bulk (Section~\ref{sec:pos_recon}). These have been performed within and beyond our WIMP search region, down to the energy threshold. Further calibration efforts are ongoing.

\subsection{Fast neutron scattering at Notre Dame}

We aim to calibrate the ionization efficiency of silicon recoils by relying on the elastic scattering of neutrons from a low energy beam in a silicon target. The pulsed beam, with a broad energy spectrum up to 600\,keV, will be produced by 2.3\,MeV protons from the Tandem Van de Graaff generator at the University of Notre Dame impinging on a \lis\ target. The experiment will be run in coincidence mode between a 30\,mg silicon drift diode and plastic scintillator detectors. The time-of-flight information between the pulsed beam, the signals in the silicon detector, and the signals in the scintillator detectors will provide discrimination from prompt $\gamma$-rays from the beam and will allow us to determine the incident neutron energy. The geometry of the coincidence setup will uniquely determine the energy of the observed silicon recoils. A similar experiment to calibrate for recoils in liquid argon has been carried out by the SCENE Collaboration~\cite{Alexander:2013aaq}. Our goal is to calibrate the ionization efficiency in silicon down to 1\,k\evr. A preliminary run with only scintillator detectors took place in March, 2013. The first run with a silicon detector is scheduled for November, 2013.

\subsection{Activated EC isotopes in the CCD}

In September, 2013 we irradiated a DAMIC CCD with a flux of \powert{10}{2} 230\,MeV protons/cm$^{2}$ at the Warrenville proton beam facility. The instrumental performance is as expected from previous radiation tolerance measurements~\cite{Dawson:2007yi}. The aim of this irradiation was to produce uniformly distributed \ber\ and \sodium\ within the CCD bulk. These isotopes decay by electron-capture (EC) and, as the $\nu$s and $\gamma$-rays escape the CCD, the only energy deposited is that from the refilling of the K-shell vacancy, leading to mono-energetic deposits of nominally 54\,\eve\ and 849\,\eve. A small energy shift due to the energy deposited by the recoiling nucleus following $\nu$ and $\gamma$-ray emission is also expected. Furthermore, the total activation of these isotopes can be measured precisely from the emitted $\gamma$-rays with a Ge detector. These lines will allow us to further characterize the detector to sub-k\eve\ energy deposits in the bulk, and to demonstrate the detection efficiency of the CCD for low energy events near our threshold.

\subsection{Nuclear recoil energy calibration with a thermal neutron source}

We are pursuing the calibration of the ionization efficiency of nuclear recoils in Si at  $\sim$1\,k\evr, crucial in understanding the energy spectrum of a potential WIMP signal in DAMIC100. The strategy is to expose a Si detector to a flux of thermal neutrons and rely on the reaction~\cite{Raman:1992zz}:

\begin{equation}
^{A}\mbox{Si} + n \longrightarrow\ ^{A+1}\mbox{Si} + \gamma\mbox{s}
\end{equation}

where the $^{A+1}$Si nucleus recoils from the $\gamma$-ray emission due to momentum conservation. If only one $\gamma$-ray is emitted, or the lifetime of nuclear states in the $\gamma$-ray cascade is much greater than the stopping time of the recoils, then the total nuclear energy deposit is mono-energetic. Considering the maximum $\gamma$-ray energy of $\sim$10\,MeV, these recoil lines have energies $<$2\,k\evr. If a Si detector is exposed to a thermal neutron beam, and the coinciding $\gamma$-rays from the interaction are detected in a secondary detector, then the nuclear recoils can be effectively tagged by a time coincidence. As the recoil energy is known from the kinematics of the reaction, the nuclear recoil ionization efficiency can be measured.

As good time resolution is required to observe the coincidence, a CCD cannot be used for this calibration. We have already attempted this measurement by exposing a LAAPD~\cite{Pansart:1997ci} to a thermal neutron beam in the LENS facility at Indiana University and in a research reactor at Ohio State University, with negative results in both cases. The instrumental integrity of the LAAPD could not withstand the large neutron flux and associated backgrounds. We plan to attempt this measurement  in the near future with a Si-Li detector.

\section{Conclusion}
\label{sec:conclusion}

The DAMIC collaboration has characterized and demonstrated the potential of CCDs as low-threshold particle detectors for rare event searches. A set of these detectors have already been deployed and operated successfully at SNOLAB. An unexpected radioactive background from uranium decays in the CCD support has been identified and resolved. A limiting background at  \powert{2}{5}\,\dru\ remains. Thicker, more massive detectors in a lower-radioactivity package will be deployed in the near-future, with the goal of demonstrating the stability and background necessary for DAMIC100. DAMIC100 will consist of eighteen 5.5\,g CCDs (100\,g total) to be deployed in SNOLAB in 2014. Considering the expected operating threshold of 100\,\eve\  and limiting radioactive background of 0.5\,\dru, DAMIC100 should be able to directly test the CDMS-Si signal within a year of operation.

\section{Acknowledgements}
\label{sec:acknowledgements}

We are grateful to the following agencies and organizations for financial support: Kavli Institute for Cosmological Physics at the University of Chicago through grant NSF PHY-1125897 and an endowment from the Kavli Foundation, and DGAPA-UNAM through Grants PAPIIT No. IN112213 and No. IB100413. 

\appendix
\section{Status as of April, 2014}
\label{sec:addendum}

In the six months since TAUP2013, considerable progress has been made in the DAMIC program. The most note-worthy achievements in this period are briefly outlined below:

\begin{enumerate}[(i)]
\item  At the end of 2013 we carried out a two-month run with a higher dynamic range to perform $\alpha$ spectroscopy. From these data, we placed limits for the \ura\ and \tho\ contamination in the CCD bulk of $<$0.14\,mBq/kg (35\,ppt) and $<$0.22\,mBq/kg (18\,ppt), respectively.
\item The February, 2014 upgrade was successful and we are currently running with two 500\,\um\ thick and one 650\,\um\ thick CCDs at SNOLAB. The dark current has been measured to be 0.1\,$e^-$/pix/day, ten times better than that reported in Section~\ref{sec:e_threshold}. The readout noise has also improved to 1.8\,$e^-$ RMS, which corresponds to  6.5\,\eve.
\item After the replacement of the support piece with high-purity silicon the total count rate in the CCD has decreased from \powert{5}{1.6}\,\ru\ with V2 packaging (Fig.~\ref{fig:snolab_setup}) to \powert{4}{3.5}\,\ru. The lead shield upgrade, which should decrease the background by \powero{2}--\powero{3}, is scheduled for May, 2014.
\item Using 18.6\,days of data acquired with the 650\,\um\ thick CCD ($>$97\% duty cycle) we have performed the spatial coincidence search to set upper limits on the \sitwo\ and \pbten\ decay rates in the silicon of $<$4.8\,mBq/kg and $<$0.8\,mBq/kg, respectively.
\end{enumerate}





\bibliographystyle{elsarticle-num}
\bibliography{myrefs}







\end{document}